\documentclass[sigconf]{acmart}

\AtBeginDocument{%
  }

\copyrightyear{2023} 
\acmYear{2023} 
\setcopyright{rightsretained} 
\acmConference[CHI '23]{Proceedings of the 2023 CHI Conference on Human Factors in Computing Systems}{April 23--28, 2023}{Hamburg, Germany}
\acmBooktitle{Proceedings of the 2023 CHI Conference on Human Factors in Computing Systems (CHI '23), April 23--28, 2023, Hamburg, Germany}\acmDOI{10.1145/3544548.3581162}
\acmISBN{978-1-4503-9421-5/23/04}

\usepackage{subcaption}

\begin{document}


\title{Wish You Were Here: Mental and Physiological Effects of Remote Music Collaboration in Mixed Reality}

\author{Ruben Schlagowski}
\orcid{0000-0002-3516-6188}
\affiliation{%
  \institution{Chair for Human-Centered Artificial Intelligence, University of Augsburg}
  \city{Augsburg}
  \country{Germany}
}
\email{ruben.schlagowski@uni-a.de}

\author{Dariia Nazarenko}
\orcid{0000-0003-1164-7283}
\affiliation{%
  \institution{Chair for Human-Centered Artificial Intelligence, University of Augsburg}
  \city{Augsburg}
  \country{Germany}
}
\email{dariia.nazarenko@uni-a.de}

\author{Yekta Can}
\orcid{0000-0002-6614-0183}
\affiliation{%
  \institution{Chair for Human-Centered Artificial Intelligence, University of Augsburg}
  \city{Augsburg}
  \country{Germany}
}
\email{yekta.can@uni-a.de}

\author{Kunal Gupta}
\orcid{0000-0003-3963-8856}
\affiliation{%
  \institution{Empathic Computing Lab, University of Auckland}
  \city{Auckland}
  \country{New Zealand}
}
\email{kgup421@aucklanduni.ac.nz}

\author{Silvan Mertes}
\orcid{0000-0001-5230-5218}
\affiliation{%
  \institution{Chair for Human-Centered Artificial Intelligence, University of Augsburg}
  \city{Augsburg}
  \country{Germany}
}
\email{silvan.mertes@uni-a.de}

\author{Mark Billinghurst}
\orcid{0000-0003-4172-6759}
\affiliation{%
  \institution{Empathic Computing Lab, University of Auckland}
  \city{Auckland}
  \country{New Zealand}
}
\email{mark.billinghurst@auckland.ac.nz}

\author{Elisabeth André}
\orcid{0000-0002-2367-162X}
\affiliation{%
  \institution{Chair for Human-Centered Artificial Intelligence, University of Augsburg}
  \city{Augsburg}
  \country{Germany}
}
\email{elisabeth.andre@uni-a.de}

\renewcommand{\shortauthors}{Schlagowski et al.}

\begin{abstract}
With face-to-face music collaboration being severely limited during the recent pandemic, mixed reality technologies and their potential to provide musicians a feeling of "being there" with their musical partner can offer tremendous opportunities. In order to assess this potential, we conducted a laboratory study in which musicians made music together in real-time while simultaneously seeing their jamming partner's mixed reality point cloud via a head-mounted display and compared mental effects such as flow, affect, and co-presence to an audio-only baseline. In addition, we tracked the musicians' physiological signals and evaluated their features during times of self-reported flow. For users jamming in mixed reality, we observed a significant increase in co-presence. Regardless of the condition (mixed reality or audio-only), we observed an increase in positive affect after jamming remotely. Furthermore, we identified heart rate and HF/LF as promising features for classifying the flow state musicians experienced while making music together.
\end{abstract}




\maketitle

\begin{figure*}
\centering
\includegraphics[width=0.89\textwidth]{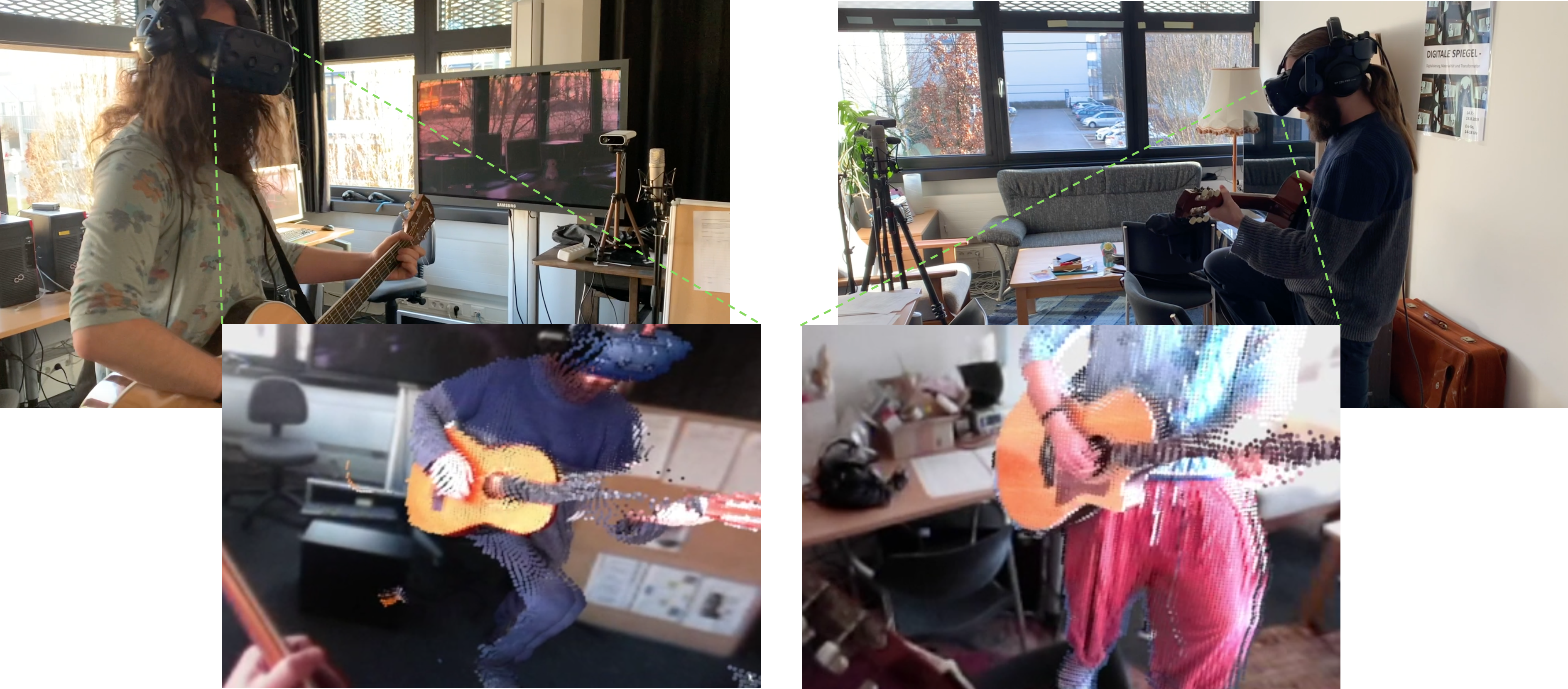}
\caption{Two jamming partners in separate rooms and their mixed reality views featuring 3D point clouds during a jam session.}
\Description{Two jamming partners playing guitar while wearing HTC Vive Pro headsets in separate rooms. Below, their mixed reality views featuring 3D point clouds during a remote jam session are shown.}
\label{fig:pointclouds}
\end{figure*}

\section{Introduction}

For many creative people, music is a medium that allows them to express themselves and connect with people. Several studies have shown that consumption and active music participation can positively affect well-being and self-esteem \cite{hallam2008effects, macdonald2013music}. During the global COVID pandemic, many musicians were deprived of the opportunity to share their music with others in persona. Worldwide, professional and amateur musicians have been massively affected by restrictions prohibiting music rehearsals, jam sessions, and live concerts. Pandemic-induced stress negatively impacted the motivation of musicians that desire external triggers to be creative or regard music-making as a social activity \cite{cai2021breakdowns}. Accordingly, artists and performers suffer from elevated anxiety, and depression \cite{burstyn2021symptoms}. In order to stay musically active in social contexts, some musicians adapted and developed asynchronous collaborative practices, mainly due to internet latency \cite{cai2021breakdowns}. However, for musical activities with real-time requirements such as improvisation or rehearsals, one common problem is that tools such as \emph{Jamulus}\footnote{\url{https://jamulus.io/}} typically occupy all available bandwidth for low-latency audio while neglecting non-auditory modalities. Consequently, critical non-auditory means of communication such as posture, gestures, or facial expressions are rarely transmitted, and the subjective experience of social presence or co-presence while jamming remains limited.

Mixed Reality (MR) technologies such as Augmented Reality (AR) and Virtual Reality (VR) can provide users with an increased feeling of presence, which was found to correlate positively with the feeling of co-presence or "being there together" with someone at the same time \cite{schroeder2002copresence, slater2000small}. Even though the idea of using MR technologies for an increased feeling of co-presence in a real-time jamming situation seems trivial, its potential has been under-researched up to this point, mainly due to latency-related technical problems. Our paper, however, does not focus on solving the latency problem but ventures into the future with potentially better bandwidths that allow low-latency transmission of high-quality audio and mixed reality content over increasing distances. We do so in order to answer the question of whether mixed reality approaches are promising regarding desirable mental effects.
Besides co-presence, one of such desirable mental effects in the jamming context would be to foster the psychological state of flow, which has been described as a delightful psychological state that refers to the "holistic sensation people feel when they act with total involvement (in an activity)" \cite{csikszentmihalyi2000beyond}. Another would be to induce a good mood or positive affect for participants, which may affect well-being in the long term \cite{diener1984independence}.

In this paper, we examine the impact of a jamming experience that makes it seem as if the remote jamming partner is "there" in the same room on co-presence, flow, and positive affect. To this end, we conducted a laboratory study featuring pairs of musicians jamming with each other in separate rooms. Following the procedure of a within-subjects experiment, we looked for differences in the aforementioned mental effects for musicians that either jammed in a audio-only setup, or while seeing their musical partners' mixed reality point clouds via a head-mounted display. Questionnaire analysis revealed that musical improvisation in a remote, low-latency situation significantly increased participants' positive affect, regardless of whether they were in the audio-only or mixed reality condition. Further, we found a significant increase in the feeling of co-presence in the mixed reality condition. However, we did not observe any significant difference regarding positive affect or flow between both conditions. 

In order to assess the feasibility of an affective mixed reality system that stimulates or assists users to get more easily into the state of flow, we tracked the musicians' Heart Rate Variability (HRV) and Electrodermal Activity (EDA) signals while jamming and evaluated their frequency and time domain features. For time intervals during which participants self-reportedly experienced flow in mixed reality, we observed strong correlations with heart rate measured in BPM and the ratio of low and high frequency components (LF/HF). These findings encourage further development and research regarding affective computing methods incorporating real-time flow assessment using physiological signals.

To contextualize and discuss the observed effects and to be able to design improved mixed reality systems for real-time music collaboration, we asked participants for qualitative free-form feedback, which we analyzed using inductive thematic analysis \cite{braun2012thematic}.

\section{Related Work}
\label{sec: related work}

\subsection{Mental Effects of Music Collaboration}
\label{subsec: Mental Effects of Music Collaboration}

Numerous studies have shown the beneficial effects of music-related activities on people of different ages, cultural backgrounds, and levels of musical affinity \cite{welch2020impact}. The benefits range from raising one's self-esteem in the short term \cite{elvers2018music} to improving an individual's mood \cite{schafer2020music} and cognitive abilities \cite{schellenberg2005music}. Keeler et al. found that group singing reduces stress and arousal and that plasma oxytocin levels in musicians' blood after improvisational singing were raised, which could "be attributed to the social effects of improvising musically with others" \cite{keeler2015neurochemistry}. 

Furthermore, musicians that participate in music-making activities can enter the state of flow \cite{forbes2021giving, de2010psychophysiology}, which was originally introduced by Csikszentmihalyi et al. in 1975 \cite{csikszentmihalyi1975beyond}. In the musical context, this state can be described as a joyful state of complete absorption in the act of music-making \cite{csikszentmihalyi2000beyond}. While flow is unlikely to be directly taught or induced, improving the conditions for the flow state may increase its likelihood \cite{jackson1999flow}. For musicians, this may be achieved by providing means that enable optimal performance, such as interpersonal connectedness, emotional connectedness, or a suiting environmental context \cite{ford2020pouring}. Overall, flow research in a musical context has been gaining momentum recently, with the most typical assessment tools being quantitative self-report questionnaires and qualitative interviews \cite{tan2021flow}. In 2007, Fritz et al. observed that musicians could experience flow in various musical activities and that the occurrence of flow correlates with well-being \cite{fritz2007experience}. In 2023, Wrigley and colleagues found that flow experience in musical practice did not fluctuate substantially for different age groups, genders, or used music instrument families \cite{wrigley2013experience}. In accordance with these findings, the state of flow has been observed with improvising jazz musicians \cite{forbes2021giving} and piano players \cite{de2010psychophysiology}. Loepthien et al. discovered that flow and subjective well-being are related for individuals with a highly flexible self-concept and evidence between flow, previous musical practice, and music experience \cite{loepthien2022flow}. It should be noted that well-being as a term is often used synonymously with positive affect, which, according to multiple studies, can be increased through interaction with music \cite{fredenburg2014effects, merry2021effects, dunbar2012performance, groarke2019listening}. Subjective well-being, however, was shown to be comprised of both positive and negative affect \cite{diener1984independence}, which have a complex relationship with one's long-term well-being and happiness \cite{diener1984independence}.

\subsection{Music and Remote Collaboration}
\label{subsec:Music and Remote Collaboration}
There exists plenty of work that tackles remote collaboration in a musical context.
One of the most critical requirements for music collaboration software is low latency \cite{schlagowski2022jamming}. 
Previous systems put much effort into bypassing that problem, e.g., by communicating not directly in the audio domain but with intermediate concepts like collaborative synthesizer-coding \cite{de2015supercopair, oh2010evolving} or directly limiting the system to work with few, predefined sounds, like the online-jamming tool \emph{Plink}\footnote{\url{https://plink.in/}} does. 
Other approaches directly address the problem of low latency. For example, \emph{Jamulus}\footnote{\url{https://jamulus.io/}} and \emph{Jacktrip}\footnote{\url{https://jacktrip.org/}} offer the possibility to make music together in real-time, and they have proven to work well if configured properly and executed on hardware that is powerful enough \cite{mall2021low}. 
While those two allow musical collaboration with direct audio signals, other systems focus on specific instruments like electronic sequencers \cite{carot2009netjack}, or synth instruments \cite{gurevich2006jamspace}. 
Other concepts focus more on collaborative idea sharing via visual music representations \cite{bryan2004daisyphone1, bryan2004daisyphone2}.

However, it is crucial for musicians to use instruments they are proficient with and not to be restricted to a particular, predefined type of instrument, even for remote jamming \cite{schlagowski2022jamming}.
There have been some efforts in the research field of \textit{Networked Music Performance} \cite{rottondi2016overview} to include visual modalities for remote music collaboration, such as the LOLA system by Drioli and colleagues \cite{drioli2013networked}. Unfortunately, such approaches were rarely continued, even though it was found that video streams can increase effectiveness in situations such as remote music lessons \cite{davies2015effectiveness}. Commercial tools like Jamulus or Jacktrip follow this trend and ignore visual modalities. Consequently, some musicians and music teachers even use video-conferencing tools like \emph{Zoom}\footnote{\url{https://zoom.us/}}\cite{macdonald2021flattening, leicht2022teaching} for musical collaboration, which has flaws regarding audio latency \cite{cai2021breakdowns}. 

To our knowledge, no commercially available application exists that allows playing with any musical instrument of choice in both the audio and visual domain while keeping audio latency at a minimum in a collaborative setting. In order to research the effectiveness of a mixed reality based approach to incorporate visual modalities for networked music performance, we used a lab prototype developed by Schlagowski et al. \cite{schlagowski2022jamming} that uses analog audio transmission and point clouds that are transmitted via a local area network to meet these requirements.

\subsection{Embodied Avatars and Social Presence in MR/XR}
\label{subsec:embodieda_vatars}

One key consideration when designing social mixed reality systems is the representation of other users in shared virtual or augmented environments. Early work on avatars in virtual environments did find evidence that interaction with a less anthropomorphic avatar led to increased social or shared presence, but this effect was attributed to expectations of anthropomorphism that could not be met at the time  \cite{nowak2003effect}. These findings are heavily contrasted by later work featuring improved technical setups, which found that high-fidelity capture and display of movement for embodied avatars can strongly increase the feeling of social presence \cite{greenwald2017investigating}. For remote collaboration tasks in mixed reality, providing gesture cues alongside gaze cues \cite{bai2020user} or even miniature 3D avatars of collaborative partners \cite{piumsomboon2018mini} can significantly increase the sense of co-presence. This is backed by the findings of Smith et al., who found that embodied avatars in VR can provide high levels of social presence and even result in similar conversation patterns to face-to-face interaction \cite{smith2018communication}. On the other hand, providing only the shared environment was perceived as lonely and led to reduced communication \cite{smith2018communication}.

While the studies mentioned above used embodied avatars that are comprised of pre-designed 3D meshes, a greater step towards anthropomorphism and photorealism is the use of depth cameras to recreate a hologram of a remote person in real-time, which makes it seem as if a person is teleported into the surroundings of a remote user \cite{orts2016holoportation}. Early studies show that this approach can make the interaction feel more natural compared to video conversations \cite{orts2016holoportation}. In line with this work, which highly encourages photorealistic and motion-tracked avatars to maximize co-presence and social presence, we used an approach similar to Orts et al. \cite{orts2016holoportation} by tracking the depth-image of a musical partner and transmitting it to a remote setup where it was then reconstructed as a 3D point cloud hologram.

\subsection{Mixed Reality and Music Collaboration}
\label{subsec:XR and Music and Collaboration}

A handful of work can be found that incorporates Mixed Reality (MR) or Extended Reality (XR), both terms including Augmented Reality (AR) and Virtual Reality (VR), into the design of music collaboration systems. An overview of this work can be found in work by Turchet et al. \cite{turchet2021xrmusic}, where collaborative \textit{Musical XR} systems were categorized as being \textit{networked systems} or \textit{multi-user experiences}. While the former describes systems "that are capable of communicating with
a plethora of devices and enabling multi-user interactions", the latter are systems where the opportunity is provided to "create shared social experiences" \cite{turchet2021xrmusic}. Both system types can support on-site or remote collaboration \cite{turchet2021xrmusic}. 

Early collaborative musical XR systems typically used tangibles such as Fiducial markers or cards that could be manipulated on-site collaboratively \cite{poupyrev2000augmented, kusunoki2002symphony, barrass2006musical}. Later work incorporated elements such as minigames \cite{pfaff2015games} or gesture-controlled virtual instruments \cite{hamilton2016gesture} that could be used to create and manipulate sounds together. Other ideas include the collaborative use of virtual step sequencers \cite{men2018lemo}, or loopers \cite{park2019arlooper} which could be manipulated together in shared virtual or augmented spaces. However, for all the work we mention, mixed reality is typically used to either comprise or be a part of virtual instruments. To our knowledge, the focus of fostering co-presence through mixed reality systems, e.g., reconstructing a remote musician in the users surroundings in real-time, was neither implemented nor researched in a musical context.

Key design requirements for a system that enables real-time remote music collaboration in MR are that (1) musicians need to be able to see their own musical instruments while (2) simultaneously seeing their remote partner with their musical instrument of choice in real-time. 
Mixed reality solutions that could fulfill these requirements are, e.g., MR solutions incorporating 3D avatars \cite{yoon2019effect}, perspective-corrected life-size projections \cite{pejsa2016room2room}, or solutions that include 3D data capturing and reconstruction \cite{regenbrecht2019preaching, lindlbauer2018remixed}. While the first approach heavily relies on pre-installed assets such as 3D avatars (and for music collaboration 3D instrument proxies), the others only rely on capturing and reconstructing sensor data, thus not limiting instrument choice while maintaining high fidelity for posture and gesture transmission. 
Reconstruction of sensor data in the remote user environment can be achieved by rendering voxels on HMDs as encouraged by Regenbrecht et al., who also demonstrated real-time applications \cite{regenbrecht2019preaching}. This approach has also been used to playback immersive 3D videos \cite{regenbrecht2021voxelvideos} or for interacting with reconstructed environments \cite{lindlbauer2018remixed}. Schlagowski et al. proposed a similar reconstruction of 3D sensor data as an overlay for pass-through AR based on their findings from interviewing musicians in focus groups \cite{schlagowski2022jamming}.

\subsection{Physiological Signals, Music Making and Flow}
\label{sec:PhysFlow}

Previous studies by Morgan and colleagues have investigated affective and behavioral sensors \cite{morgan2015using} and physiological signals \cite{morgan2014instrumenting} during musical improvisation. In the former study, the authors reported alignments between rhythmic change points (points in time when rhythmical patterns changed) and the heart rate of jamming musicians. In the second paper, correlations of Electroencephalography (EEG) and Galvanic Skin Response (GSR) measures with self-reported levels of Energy, Positivity, and Boredom were observed \cite{morgan2014instrumenting}. These self-reported measures, however, were not psychometrically validated.
A psychological state that was widely adopted in psychology and repeatedly found to be occurring in music practice \cite{forbes2021giving, de2010psychophysiology, wrigley2013experience} is the state of flow (see subsection \ref{subsec: Mental Effects of Music Collaboration}).
For the musical context, De Manzano et al. proposed that "flow is the subjective experience of an interaction between positive valence and high attention during the performance" \cite{de2010psychophysiology}. This notion of underlying physiological mechanisms has been demonstrated through empirical results suggesting flow state's association with Autonomic Nervous System \cite{peifer2014relation}, cortex activity \cite{dietrich2004neurocognitive}, and dopamine \cite{davis2009dopamine}. Peifer et al. suggested that an inverted u-curve nonlinear function is followed in the relation of flow with sympathetic arousal and hypothalamic-pituitary-adrenal (HPA) axis activation, i.e., flow experience should be facilitated by moderate physiological arousal \cite{peifer2014relation}. In contrast, the flow state should be hindered by excessive physiological arousal. Through their research, Harmat et al. \cite{harmat2015physiological} demonstrated an association of flow state with respiratory depth and Low-Frequency Heart Rate Variability (LF-HRV). 
Examining the flow condition in VR games, Tozman et al. reported that flow experience was associated with lower LF-HRV activity in a balanced skill task \cite{tozman2015understanding}.

Skin Conductance or Electrodermal Activity (EDA) is an indicator of general arousal and attention and acts as a reliable measure of sympathetic activation. Kivikangas et al. \cite{kivikangas2006psychophysiology} demonstrated that flow could be investigated through EDA measurements as flow state is usually associated with physiological arousal, which can be reliably reflected through EDA \cite{engeser2012advances}. Higher flow levels are related to moderate EDA levels, which can be visualized in an inverted-u function. Bian et al. \cite{bian2016framework} provided a comprehensive evaluation model for measuring flow experience in VR games using physiological signals such as Heart Rate Variability, Respiration, and facial Electromyography (EMG). These past researches guided us to select EDA and HRV as primary physiological signals to measure and quantify the flow state. The raw signal we measured for EDA is Gavanic Skin Response (GSR) and for HRV we measured Photoplethysmogram (PPG).

\section{Methodology}
\label{sec: Methodology}

We conducted a laboratory study, including a within-subjects experiment between February and May 2022 in Germany. During this experiment, we let participants improvise in pairs of two with their instruments of choice. We measured and compared the psychological effects of jamming in an \textit{AudioOnly} and \textit{MixedReality} condition. The following subsection describes these conditions and reflects on our motivation and reasoning behind the condition choice.

\begin{figure*}[h!]
\centering
\includegraphics[width=0.80\linewidth]{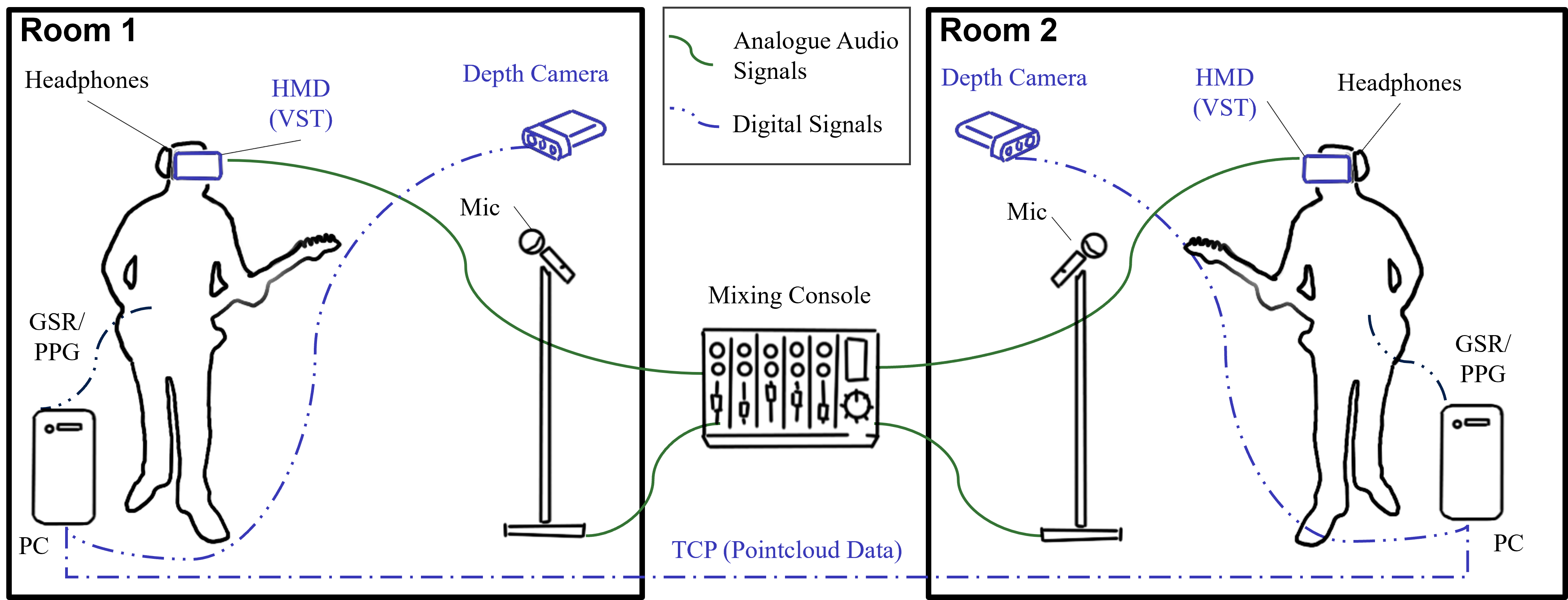}
\caption{The experiment setup. Blue elements were only used in the \textit{MixedReality} condition.}
\Description{A schematic of the technical setup is displayed, showing technical components that are installed in two separate rooms. Both rooms are equipped with a microphone and a depth camera. The microphones are connected to a mixing console that is placed outside both rooms. Participants wear PPG/GSR sensors that are connected to a PC and wear headphones that are connected to the mixing console. Both PCs are connected via TCP that transmits point cloud data.}
\label{fig:setup}
\end{figure*}

\subsection{Baseline Choice and Condition Design}
\label{sec: ConditionDesign}

For our \textit{MixedReality} (MR) condition, we chose the approach by Schlagowski et al. \cite{schlagowski2022jamming}, as it was designed based on the findings of focus group interviews with musicians and reportedly met with positive feedback during early tests. 
Using this approach, musicians can see their own instruments via the video pass-through of mixed reality headsets and their partner musician's reconstruction along their physical instruments as stereoscopically rendered 3D point clouds (comprised of hexagonal shapes placed in 3D space). Compared to perspective-corrected wall projections as in \cite{pejsa2016room2room} (or even conventional 2D screens), this HMD-based approach comes with some technical limitations, especially regarding video stream and point cloud resolution. However, by using commercially widely available head-mounted displays, we ensure both a low-cost reproducibility for the research community and a condition design that resembles the de facto state-of-the-art for hands-free mixed reality experiences for a larger user group.

To rule out audio-related issues and inconsistencies as potential disturbance factors between conditions, we used analog audio hardware instead of digital transmission protocols for audio transmission between the isolated musicians. By doing so, we maintained an unnoticeable and constant low audio latency for both conditions (in analog equipment used over short distances, latency is generally not considered a factor since the signals travel at 66\%-95\% of the speed of light, primarily dependent on the insulating material \cite{kaiser2005transmission}). As musicians need to communicate freely, e.g., to agree on a key or mode for improvisation, we used two high-sensitivity microphones for the low-latency audio transmission, capturing the whole room, including voices.

As we wanted to investigate the unaltered influence of our selected HMD-based MR approach, we designed our second condition, \textit{AudioOnly} (AO), to be a remote jamming setup that is as close to the MR condition as possible after the removal of all MR-related aspects. As such, we used the same rooms, sensors, and audio setups as in condition MR and solely removed the HMDs that rendered the real-time point clouds. By deciding against the use of HMDs in pass-through mode in the AO condition, we ensured to observe the objective impact of both positive and negative aspects that come with the selected HMD-based approach for condition MR (e.g., ergonomic issues or limited resolution). 

We want to note that the AO condition, to a limited extent, can be interpreted as a best-case for the current de facto state-of-the-art approach for networked music performance, which is latency-optimized audio-only transmission in solutions like \textit{Jamulus}. However, even cutting-edge applications may never reach a similar quality or latency over arbitrary distances (please refer to the discussion section for details). To summarize, both conditions are listed below:

\begin{itemize} 
    \item \textit{AudioOnly (AO):} Participants are situated in separate rooms while receiving high-quality and low-latency audio signals from each other (both for voices and their instruments).\\
    \item \textit{MixedReality (MR):} Participants are in the same situation as in AO but simultaneously experience the jamming partner's mixed reality point cloud in real-time via a head-mounted display.
\end{itemize}

\subsection{Technical Setup}
\label{subsec: Experiment Setup}

We prepared two separate rooms with hybrid Mixed Reality and audio setups, consisting (per room) of an HTC Vive Pro, a high-performance PC (Intel(R) Core(TM) i9-9900K CPU, 32 GB RAM, NVIDIA GeForce RTX 2080 Ti GPU), one Azure Kinect depth camera, a high-quality room microphone (Rode NT-1A), and a bodypack/headphone setup for audio (see Figure \ref{fig:setup}). Musicians could either use the pre-installed high-sensitivity room microphones or optional direct signal paths to the console (e.g., if instruments were electric and not acoustic) for audio transmission. If they chose the latter, the room microphones remained active in parallel to enable verbal communication between both rooms.

Furthermore, we let jamming partners wear \emph{Shimmer}\footnote{\url{https://shimmersensing.com/}} sensors, measuring Photoplethysmogram (PPG) for Heart Rate Variability (HRV) and Galvanic Skin Response (GSR) for Electrodermal Activity (EDA). Both signals were synchronized with the individual audio channels of the jamming partners using the SSI software \cite{wagner2013social}. All audio and physiological signals were stored on a hard disc for later evaluation and annotation. The test supervisor mixed the analog high-quality audio signals for sending and monitoring returns in a separate control room, similar to live-recording setups in professional recording studios.
The point-cloud data from the depth cameras was transmitted to the other PC via TCP sockets over the local network and visualized with a shader (a modified version of Kejiro's PCX package\footnote{\url{https://github.com/keijiro/Pcx/}}) on the VR-HMDs using the Unity game engine\footnote{\url{https://unity.com/}}. 
As these point clouds were rendered above stereoscopic real-time video streams of the participants' surroundings using the HTC SRWorks SDK\footnote{\url{https://developer.vive.com/resources/vive-sense/srworks-sdk/}}, the MR setup could be classified as a video-see-through (VST) augmented reality (AR) application. The source-code of our modified demonstrator is available on GitHub.\footnote{\url{https://github.com/hcmlab/WishYouWereHere/}}

\subsection{Participants}
\label{subsec: Participants}

We acquired a total of 26 participants (13 pairs) from Germany through several methods, including but not limited to mailing lists, social media, classified advertisements, handwritten advertisements on local notice boards, and forum posts on university-related web resources. Twenty of them identified as male, and six identified as female. The participants were aged between 18 and 69 (\textit{M}: 40.04, \textit{SD}: 16.43). Participants provided written free-form information about their musical background and expertise before the study. Participants provided written free-form information about their musical background and expertise before the study. Based on this information, eight musicians were categorized as advanced, fourteen as intermediate, and four as beginner musicians through authors' majority vote. Fourteen participants played the guitar, four bass guitar, and three the ukulele. One participant each played the accordion, mandolin, violin, vocals, or drums.

Participants were free to choose their jamming partners prior to the experiment and often chose to participate with their band/orchestra mates, colleagues, or friends. We arranged jamming partners for participants who did not form pairs on their behalf by selecting the potentially best match in terms of personal preferences and abilities from the currently available pool of other musicians without jamming partners. We handed such "freshly created" pairs their partners' contact data so they could get to know each other and discuss their musical skills/preferences in advance.  

\begin{figure}[h!]
\begin{subfigure}{.49\linewidth}
  \centering
  \includegraphics[width=0.99\linewidth]{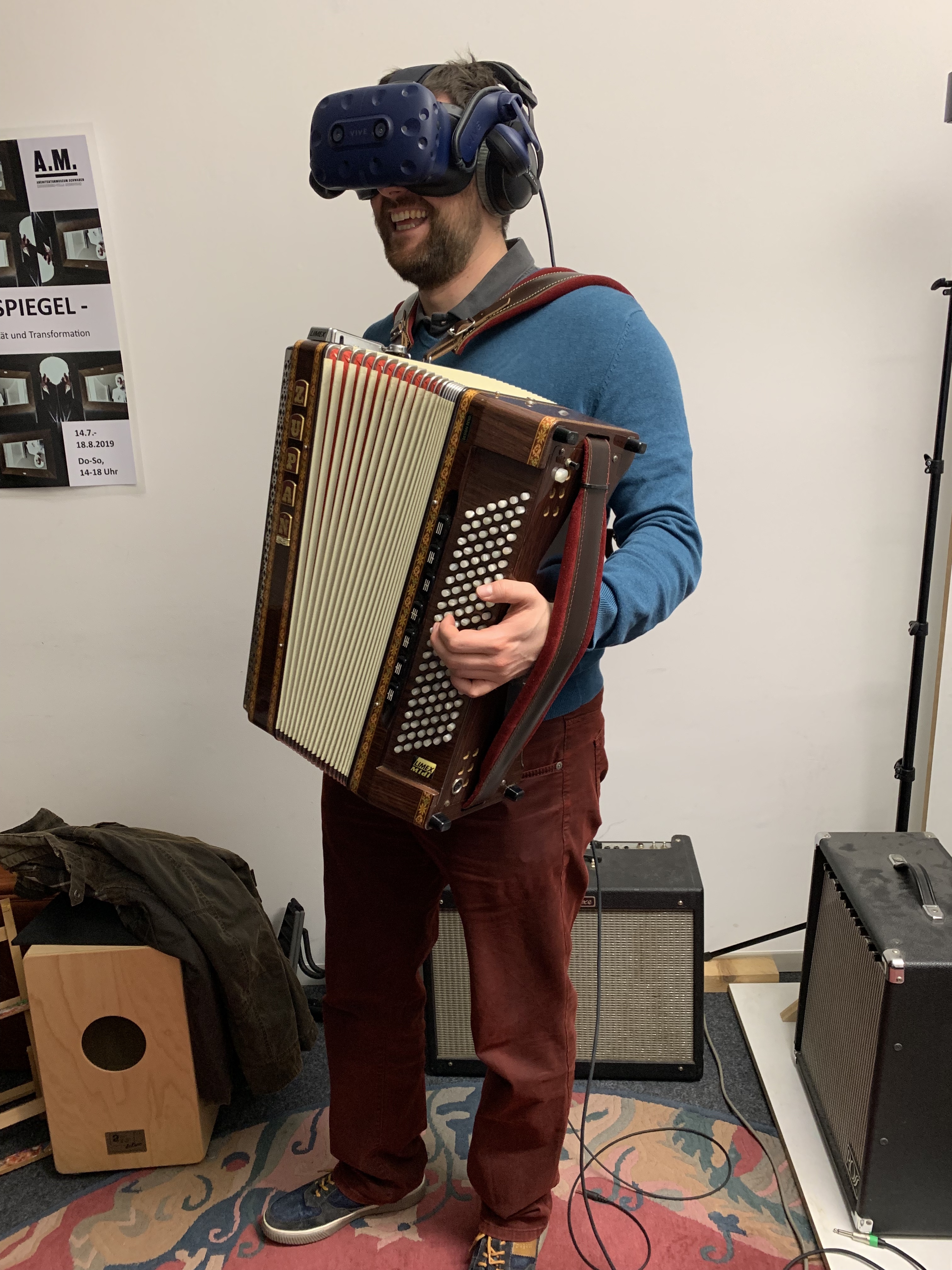}

\end{subfigure}%
\begin{subfigure}{.49\linewidth}
  \centering
  \includegraphics[width=0.99\linewidth]{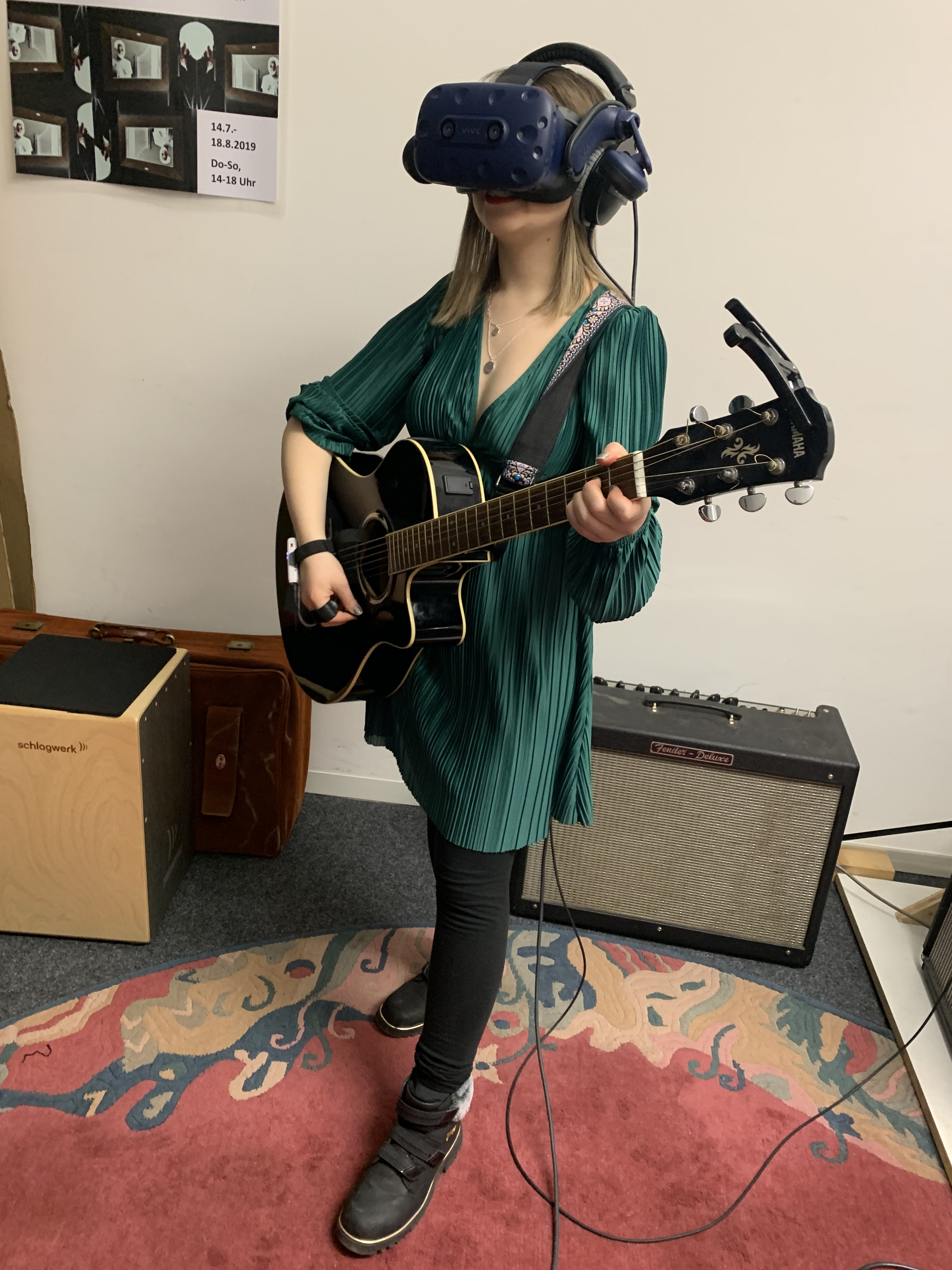}

\end{subfigure}
\begin{subfigure}{.49\linewidth}
  \centering
  \includegraphics[angle=270,origin=c,width=.99\linewidth]{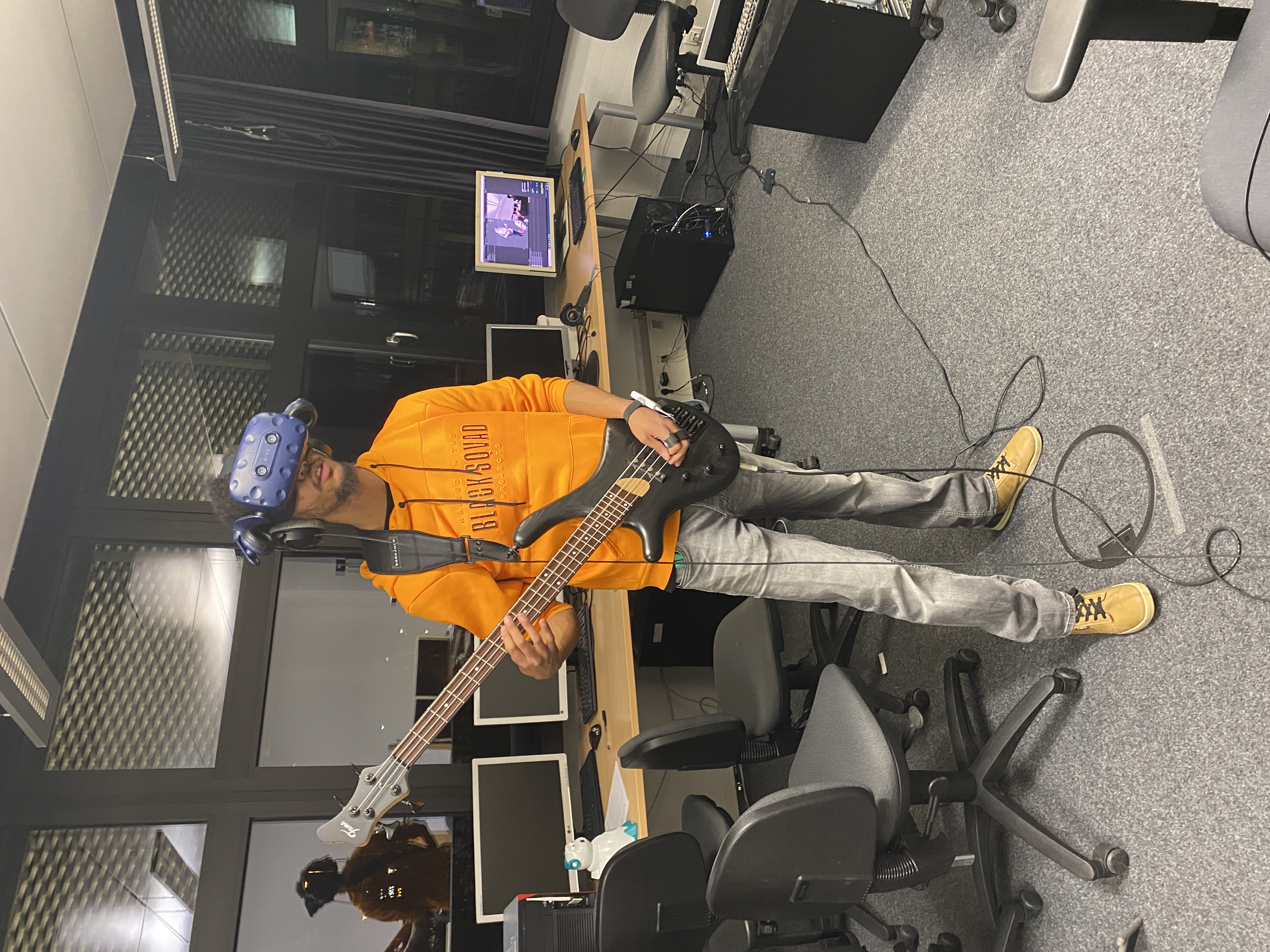}

\end{subfigure}%
\begin{subfigure}{.49\linewidth}
  \centering
  \includegraphics[angle=270,origin=c,width=.99\linewidth]{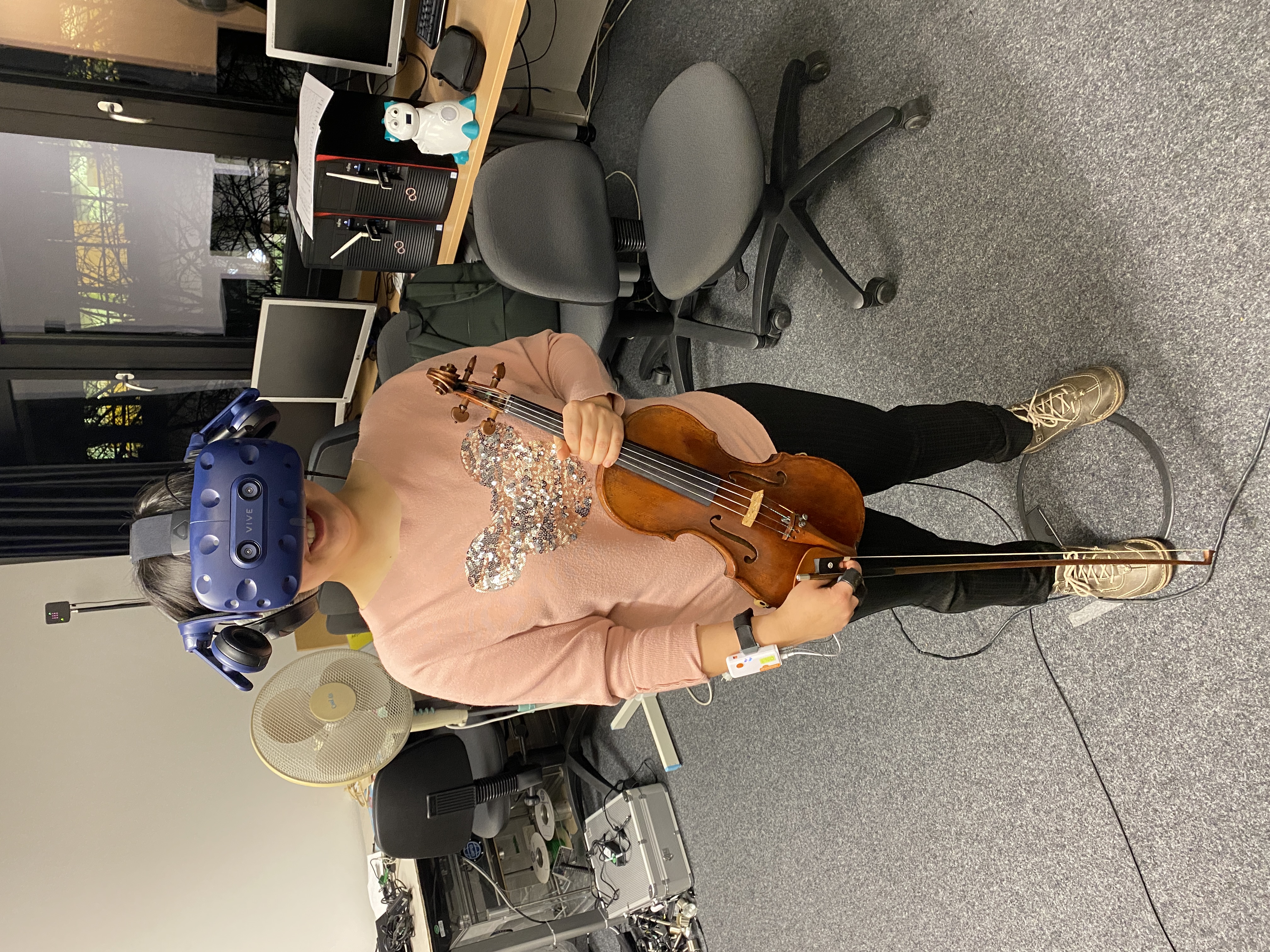}

\end{subfigure}
\caption{Four study participants with their musical instruments, jamming in the \textit{MixedReality} condition.}
\Description{Four study participants can be seen, each wearing mixed reality headsets. One plays the accordion, and one the guitar. Another is playing the bass guitar, and a fourth is holding a violin.}
\label{fig:study_pics}
\end{figure}


\subsection{Research Questions, Dependent Variables and Hypotheses}
\label{subsec: RQHypothesis}

The research questions that we address in our study are as follows:
\begin{itemize} 
  \item [RQ1:] Is the feeling of co-presence stronger for participants experiencing the real-time mixed reality point clouds of each other (condition \textit{MixedReality}) as opposed to just hearing each other (condition \textit{AudioOnly})?
  \item [RQ2:] Are the self-reported flow levels higher for jam-session in condition \textit{MixedReality} than for condition \textit{AudioOnly}? 
  \item [RQ3:] Is self-reported positive affect higher after a jam-session in condition \textit{MixedReality} than in condition \textit{AudioOnly}?
  \item [RQ4:] Does an individual's level of positive affect rise after jamming remotely with a partner (independently of both conditions)?
  \item [RQ5:] Can we observe differences in features of EDA and HRV signals comparing flow and non-flow states of users jamming in MR?
  \item [RQ6:] How does the experience of jamming in the \textit{MixedReality} condition compare to a jam in the \textit{AudioOnly} condition on a qualitative level?
  \item [RQ7:] What are the positives and negatives of the used MR system as-is, and how can it be improved? 
\end{itemize}

We addressed RQ6 and RQ7 by conducting an inductive thematic analysis \cite{braun2012thematic} on written, open feedback that participants gave after the study. The results are discussed in subsection \ref{subsec:qualitativeresults}. For RQ5, we used annotations on flow state participants made after each jam session (explained in further detail in subsection \ref{subsec: StudyProcedure}) and synchronized them with the physiological signals that we gathered during the jam sessions. Then, we explored the data to find descriptive features for the flow state (more details in subsection \ref{methoddologyphysiological}). The results of this exploratory analysis can be found in subsection \ref{subsec:physiologicalresults}. RQ1-RQ4 were addressed through statistical hypothesis testing, examining data gathered from questionnaires before, during, or after the jam sessions. The corresponding hypotheses are:

\begin{itemize} 
  \item [H1:] The self-reported feeling of co-presence after jamming in condition \textit{MixedReality} is greater than in condition \textit{AudioOnly}.
  \item [H2:] The self-reported flow level after jamming in condition \textit{MixedReality} is greater than in condition \textit{AudioOnly}.
  \item [H3:] The self-reported positive affect after jamming in condition \textit{MixedReality} is greater than in condition \textit{AudioOnly}.
  \item [H4:] The self-reported positive affect after jamming is greater than prior to jamming in a low-latency remote setup.

\end{itemize}

Co-presence was measured using the \textit{Networked Minds Social Presence Inventory} (NMSPI) questionnaire \cite{biocca2003networked}, which measures co-presence as one sub-scale amongst other dimensions of social presence. Flow was measured using the \textit{Flow Short Scale} (FSS) \cite{Rheinberg2003}, and positive affect was measured using the \textit{Positive and Negative Affect Schedule} (PANAS) \cite{tellegen1988development}. Table \ref{table:variables} summarizes the dependent variables that are statistically tested and the questionnaires used for measurement. The quantitative results of the statistical hypothesis tests can be found in subsection \ref{subsec:dependentresults}, where we plotted additional dimensions and sub-scales that the questionnaires could measure as supplementary results in Figure \ref{fig:quant_results}. However, we did not consider them dependent variables for our experiment.

\setlength{\arrayrulewidth}{0.5mm}
\begin{table}[h]
    \begin{tabular}{|p{0.4cm}|p{1.15cm}|p{3.1cm}|p{0.8cm}|p{1.1cm}|}
    \hline
    \textbf{RQ} & \textbf{Variable} & \textbf{Measured with} & \textbf{Range} & \textbf{Hypoth.}\\ 
    \hline
    1 & Co-Presence & Networked Minds Social Presence Inventory (NMSPI) \cite{biocca2003networked} & 1-7 & H1 \\
    \hline
    2 & Flow Level & Flow Short Scale (FSS) \cite{Rheinberg2003} & 1-7 & H2 \\
    \hline
    3 & Positive Affect &  Positive and Negative Affect Schedule (PANAS) \cite{tellegen1988development} & 1-7 & H3, H4 \\
    \hline
    
    \end{tabular}
    \caption{An overview of dependent variables, their respective measuring tools and hypotheses.}
    \label{table:variables}
\end{table}

\subsection{Study Procedure}
\label{subsec: StudyProcedure}

\begin{figure*}[h!]
    \centering
    \includegraphics[scale=1.0, width=0.9\textwidth]{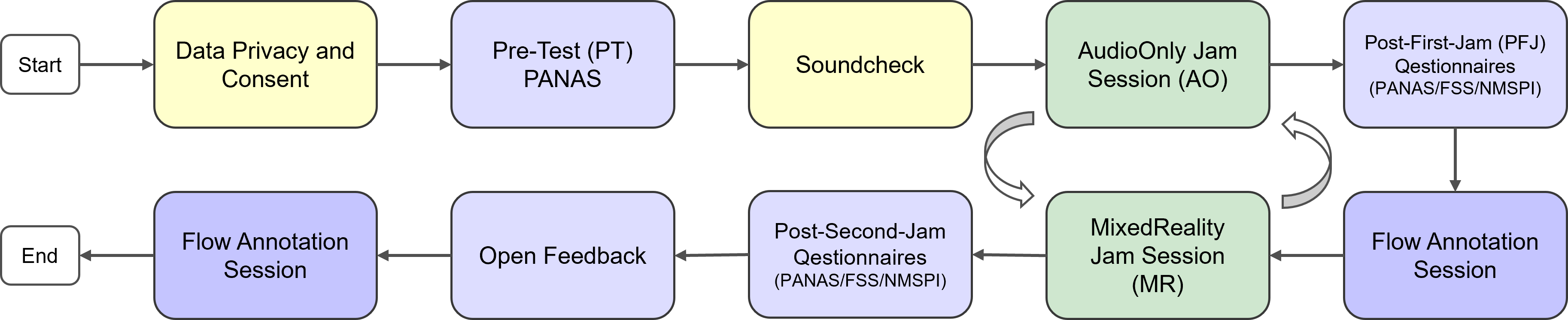}
    \caption{The experiment procedure.}
    \Description{A flow chart that visualizes the study procedure is depicted. A verbal description of the procedure can be found in subsection 3.5.}
    \label{fig:study}
\end{figure*}

As our conditions are reversible (they can be interchanged) and as our participants varied greatly regarding expertise or their preferred musical instruments, we chose a within-subjects study design (cf. Figure \ref{fig:study}). Doing so ensured a maximum measuring sensitivity for the observed differences in the dependent variables.
Thus, each participant experienced both conditions in two jam sessions of 20 minutes duration with their partner, one audio-only jam (condition \textit{AudioOnly}) and another jam featuring real-time transmissions and HMD-based visualization of mixed reality point clouds (condition \textit{MixedReality}). The order in which the participants saw these conditions was randomized for each pair of jamming partners. Thus, both partners experienced the same condition in the same time, and no mixed condition setups (\textit{MixedReality} and \textit{AudioOnly} simultaneously) did occur.

At the beginning of each experiment session, participants filled out data privacy and consent and the pre-test PANAS questionnaire to assess their positive affect before jamming. Subsequently, the test leader provided both participants with basic contextual information on the study. Afterward, participants were guided into separate rooms, and their audio setup was installed and tested. The next phase was a sound check, during which the participants could utter wishes regarding their individual monitoring mixes. As soon as both participants were content with their mix and after some further condition-specific preparations (this includes the MR system setup in condition \textit{MixedReality}), participants were given the instruction to freely improvise and the first jam session took place. Then, participants filled out post-stimulus questionnaires (PANAS and Networked Minds, measuring affect and social presence) and headed on to the first flow annotation session.

During the flow annotation sessions, the jamming partners were first provided with a verbal definition of the state of flow as defined by Csikszentmihalyi et al. (the "holistic sensation people feel when they act with total involvement (in an activity)" \cite{csikszentmihalyi2000beyond}). Then, whilst being assisted by the test supervisor, they could listen through the jam session's recorded audio and mark the time intervals during which they experienced flow state using the NOVA software \cite{Baur2020} (if this happened to be the case). Each participant could annotate this individually as all audio was recorded in separate channels. 

After the flow annotation session, the participants headed on to the second jam session under the remaining condition and another round of post-stimulus questionnaires. Then, they had the opportunity to provide additional written and open feedback, which we later used for qualitative data analysis. The experiment ended with a final flow annotation session, marking flow time intervals for the second jam session.

\subsection{Exploratory Study of Physiological Effects}
\label{methoddologyphysiological}
Heart-Rate Variability (HRV), measured via Photoplethysmogram (PPG), is one of the most commonly used physiological signals for observing varying affects. However, when it is acquired from wrist-worn devices, the signal is vulnerable to body motion. Therefore, the artifacts caused by these motions should be eliminated before further analysis. We used the HeartPy library from Python for this purpose \cite{van2018heart}. First, Hampel correction is applied to the PPG signal \cite{liu2004line}. During this stage, the local median for each data point is computed, and if a data point differs from the median by more than three standard deviations, it is classified as an artifact and replaced with the median. Then, a non-linear quotient filter based on Poincare plots is used \cite{piskorski2005filtering}. This filter is also used in the HeartPy library, and it is a powerful tool for removing incorrect beats of the PPG signal \cite{van2018heart}.\\
\indent After cleaning the PPG signal and detecting peaks, we split the signal into 1-minute windows, which overlapped by 50\%. This window size is selected because it was reported as the minimum interval for getting meaningful features (i.e., Low Frequency component (LF), High Frequency component (HF), and LF/HF ratio) from frequency domain conversion \cite{van2018heart}. We extracted time domain, frequency domain, and non-linear features from these windows that were found to be the most distinctive features in the literature \cite{PicardLab2016, Gjoreski:2016:CSD:2968219.2968306, Alberdi201649}. Features from the time domain are as follows:  
\begin{itemize}
    \item Beats per minute (BPM): Heartbeats within 60 seconds, 
    \item interbeat interval (IBI): The time interval between two consecutive beats, 
    \item the standard deviation of all interbeat intervals (SDNN), 
    \item the standard deviation of successive interbeat intervals \\(SDSD), 
    \item root mean square of differences between successive interbeat intervals  (RMSSD), 
    \item the proportion of interbeat intervals with successive differences above 20ms (pNN20), 
    \item the proportion of interbeat intervals with successive differences above 50ms (pNN50),
    \item and median absolute deviation of all interbeat intervals \\ (HR\_mad).
\end{itemize}

BPM is known to increase under anger, anxiety, joy, happiness, and embarrassment \cite{shu2018review}, whereas RMSSD and pNN50 have correlations with increased Parasympathetic Nervous System (PNS) activity which occurs when we conserve energy and engage in tend-and-befriend situations \cite{shaffer2017overview}. \\
\indent We also computed features from the frequency domain. However, since the heart peaks are not equidistant, Fast Fourier Transform (FFT) can not be used directly. One technique to overcome this issue is applying Welch's periodogram technique \cite{welch:periodogram:1967}. This method was specifically implemented for estimating the power spectral density of these types of signals. From the frequency domain, we extracted low-frequency component (LF), which is the total power spectral density in frequency spectrum between 0.05-0.15 Hz, high-frequency component (HF), which is  the total power spectral density in frequency spectrum between 0.15-0.5 Hz, and the ratio between low frequency and high frequency components (LF/HF). The LF component is assumed to be generated by the Sympathetic Nervous System (SNS), whereas the HF component is assumed to be produced by the PNS, especially under controlled situations \cite{shaffer2017overview}. A low LF/HF ratio is a sign of parasympathetic domination. On the contrary,  a  high  LF/HF  ratio suggests sympathetic domination, which occurs when confronting fight-or-flight situations. 

Non-linear features are estimated by employing different state-space domain entropy features of the HRV signal \cite{houshyarifar2017early}. We extracted four features from the non-linear domain: SD1, SD2, S and SD1/SD2. SD1 is the standard deviation of the Poincaré plot perpendicular to the identity line, while SD2 is the standard deviation of the Poincaré plot along the identity line. S is  the area of an imaginary ellipse with axes SD1 and SD2. SD1/SD2 is the division of SD1 and SD2 parameters. The SD1/SD2 ratio was found to be correlated with the LF/HF ratio \cite{shaffer2017overview}. Furthermore, we estimated breathing rate (BR) from the PPG signal by using a frequency domain approach \cite{karlen2013multiparameter}.

\section{Results}

\subsection{Hypothesis Tests (RQ1-RQ4)}
\label{subsec:dependentresults}

\begin{figure*}[h]
\begin{subfigure}{.49\textwidth}
	\centering\captionsetup{width=.90\linewidth}
	\includegraphics[width=.92\linewidth]{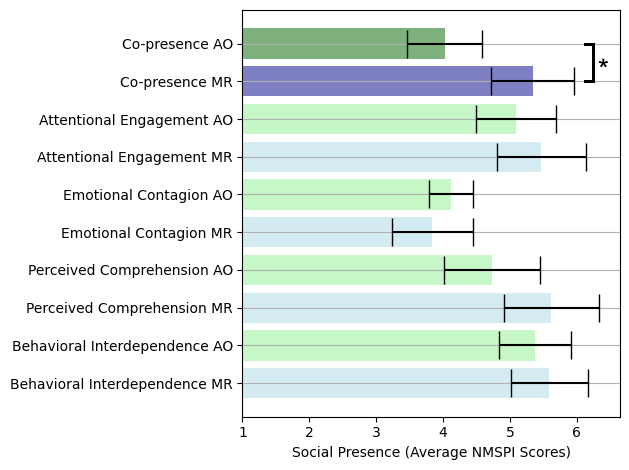}
    \caption{Social presence (NMSPI) results for conditions \textit{AudioOnly} (AO) and \textit{MixedReality} (MR).}
    \label{fig:social_presence_self_plot}
\end{subfigure}%
\begin{subfigure}{.49\textwidth}
	\centering\captionsetup{width=.90\linewidth}
	\includegraphics[width=.93\linewidth]{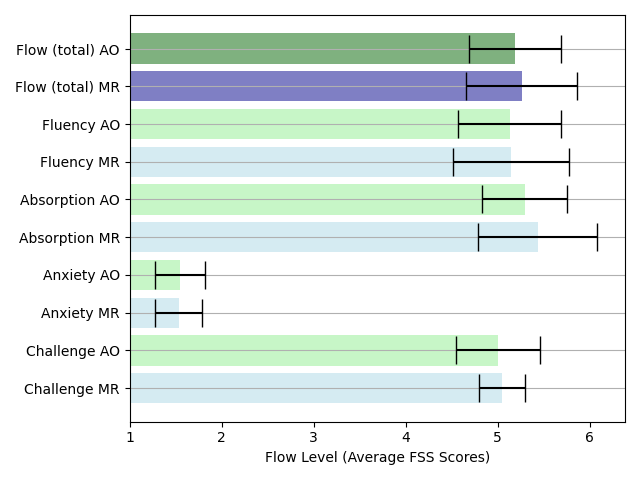}
    \caption{Flow (FSS) results for conditions \textit{AudioOnly} (MR) and \textit{Mixed-Reality} (MR).}
    \label{fig:flow_level_plot}
\end{subfigure}
\begin{subfigure}{.49\textwidth}
    \centering\captionsetup{width=.90\linewidth}
    \includegraphics[width=0.90\textwidth]{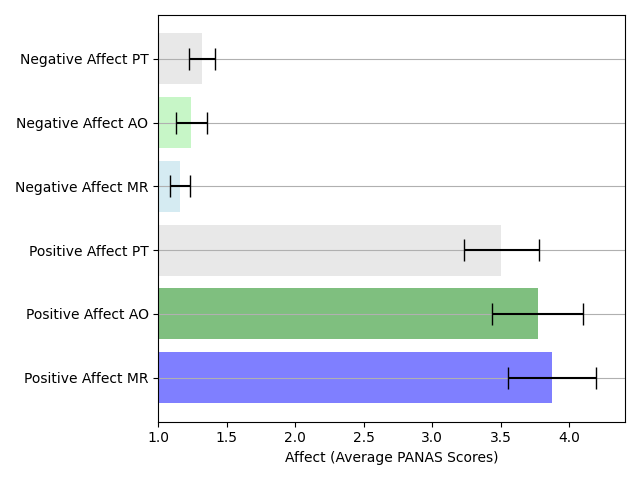}
    \caption{PANAS results for pre-test (PT) and conditions \textit{AudioOnly} (AO) and \textit{MixedReality} (MR).}
    \label{fig:affect_plot}
\end{subfigure}
\begin{subfigure}{.49\textwidth}
	\centering\captionsetup{width=.90\linewidth}
	\includegraphics[width=.90\linewidth]{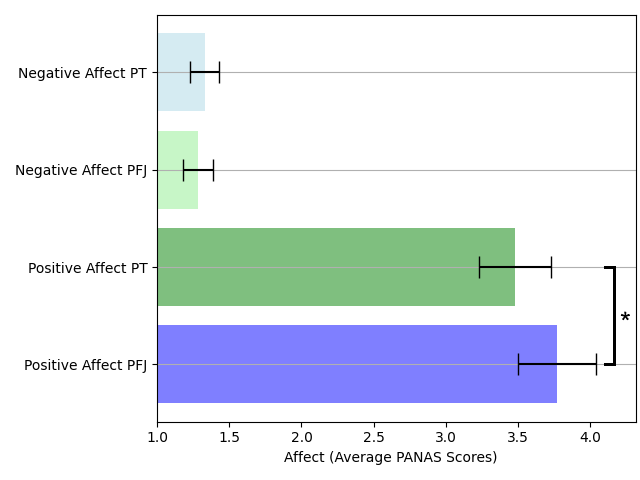}
    \caption{PANAS results for pre-test (PT) and post-first-jam (PFJ) \\affect.}
    \label{fig:affect_plot_2}
\end{subfigure}
\caption{Results of questionnaires. Bars show means and error bars show the 95.0\% confidence interval of the mean. Square brackets and asterisks indicate significant paired sample t-test results (p < 0.013). Darker bars are scores of dependent variables that were used to test hypotheses H1, H2, H3 and H4. Scores range from 1-7 for every scale.}
\Description{Four subfigures show the results of the questionnaires in bar charts. Subscales that show scores that were used for hypothesis testing are darker in color. A notable difference can be seen for co-presence between conditions A and B in subfigure 5a. Another difference can be seen in positive affect between pre-test PANAS and post-first-jam PANAS results.}
\label{fig:quant_results}
\end{figure*}

Because of technical difficulties during one experiment run, we dismissed subjective questionnaire scores (PANAS, FSS, and NMSPI) for two participants, resulting in N=24 for hypotheses H1, H2 and H4. Another participant incompletely filled out a PANAS questionnaire before the test. As this did not affect his jamming partner (it remained unnoticed), we only excluded the incomplete PANAS sample for this particular participant, resulting in a total of N=23 for hypothesis H3.

For all four hypothesis tests addressing research questions RQ1-RQ4 (see \ref{subsec: RQHypothesis}) and their corresponding hypotheses H1-H4 (see \ref{subsec: RQHypothesis}), the samples were checked for normal distribution using the Shapiro-Wilk-Test \cite{shapiro1965analysis} and for equal variances using Levene's test \cite{levene1961robust}. If the tested sample data turned out to be parametric (this requires both tests to have \textit{p}-values above the significance threshold), a paired sample t-test \cite{student1908probable} was used. If either test failed, we used Wilcoxon's signed-rank test \cite{wilcoxon1992individual}. We set the significance level to \textit{alpha} = 0.05. We applied \textit{p}-value correction using the Holm-Bonferroni method \cite{holm1979simple}.

Figure \ref{fig:qual_total} depicts results (mean values and 95\% confidence intervals) for a variety of questionnaire subscales. As we do not consider additional subscales as dependent variables, we do not report statistical analyses on them in this section. However, we refer to them in our discussion section to contextualize our findings and derive limitations. Accordingly, we interpret observations from these subscales as disputable pieces of evidence that contribute to a larger picture.

\subsubsection{Co-Presence (RQ 1)}

To test hypothesis H1 (see \ref{subsec: RQHypothesis}), we analyzed and compared the scores of the co-presence subscale of the NMSPI for conditions \textit{AudioOnly} (AO) and \textit{MixedReality} (MR). Condition MR scores were found to be normally distributed (Shapiro-Wilk test p=0.053), whereas Condition AO samples were not (p=0.026). The null hypothesis for unequal variances was rejected (Levene's test p=0.83). Consequently, the non-parametric Wilcoxon signed-rank test was used to test H1 and yielded a value of p=0.0028, indicating that there is a \textbf{statistically significant} difference between the two samples (Condition AO: \textit{M} = 4.02, \textit{SD} = 1.23; Condition MR: \textit{M} = 5.34, \textit{SD} = 1.37). This difference can also be seen in Figure \ref{fig:social_presence_self_plot}. After \textit{p}-value correction, the \textit{p}-value increased to 0.011, but the result remains significant. In \ref{fig:social_presence_self_plot}, we also present means and 95\% confidence intervals for the other four NMSPI subscales (Perceived Attentional Engagement, Perceived Emotional Contagion, Perceived Comprehension, and Perceived Behavioral Interdependence).

\subsubsection{Flow (RQ 2)}

For hypothesis H2 (see \ref{subsec: RQHypothesis}), we compared flow-scores of the Flow Short Scale (FSS) for both conditions. We found the scores for both conditions to be normally-distributed (Shapiro-Wilk test p=0.33 for condition AO, and p=0.05 for condition MR) and to have equal variances (Levene's test p=0.51). The paired sample t-test yielded a \textit{p}-value of 0.35, indicating no significant difference between the samples (Condition AO: \textit{M} = 5.19, \textit{SD} = 1.10; Condition MR: \textit{M} = 5.26, \textit{SD} = 1.32).  Figure \ref{fig:flow_level_plot} depicts both condition's mean values and standard deviations for all subscales of FSS.

\subsubsection{Positive Affect (RQ3 \& RQ4)}

We measured the levels of positive and negative affect of participants before the jam sessions (pre-test or PT) and after each condition (AO or MR) using the PANAS questionnaire \cite{biocca2001networked}. The questionnaire contains 20 items, 10 of which are used to measure positive affect, which we compared to test for hypotheses H3 and H4 (see \ref{subsec: RQHypothesis}). Negative Affect, on the other hand, was not considered a dependent variable and thus not analyzed statistically. Figure \ref{fig:affect_plot} depicts the results.

Comparing the positive affect scores in conditions AO and MR, we found the data to be normally distributed (p=0.76 for condition AO, and p=0.20 for condition MR) and to have equal variances (Levene's test p=0.54). The paired sample t-test yielded p=0.15 (Condition AO: \textit{M} = 3.77, \textit{SD} = 0.71; Condition MR: \textit{M} = 3.88, \textit{SD} = 0.69), thus rejecting H3. 

To test for H4, we compared positive affect from after the first jam (post-first-jam or PFJ), regardless of participants jammed in condition AO or MR, and compared them to the pre-test (PT) PANAS scores. The data were found to be normally distributed (p=0.71 for PT scores, p=0.48 for PFJ scores), and to have equal variances (Levene's test p=0.81). The paired samples t-test yielded p=0.0041, which implies a \textbf{statistically significant} rise in positive affect after the first jam session (PT: \textit{M} = 3.50, \textit{SD} = 0.59; PFJ: \textit{M} = 3.81, \textit{SD} = 0.62). The result remains significant after \textit{p}-value correction (adjusted p=0.012). Figure \ref{fig:affect_plot_2} depicts this result. 

\subsection{Evaluation of Physiological Data (RQ 5)}
\label{subsec:physiologicalresults}

\begin{figure*}[]
\begin{subfigure}{.49\textwidth}
	\centering
	\includegraphics[width=.99\linewidth]{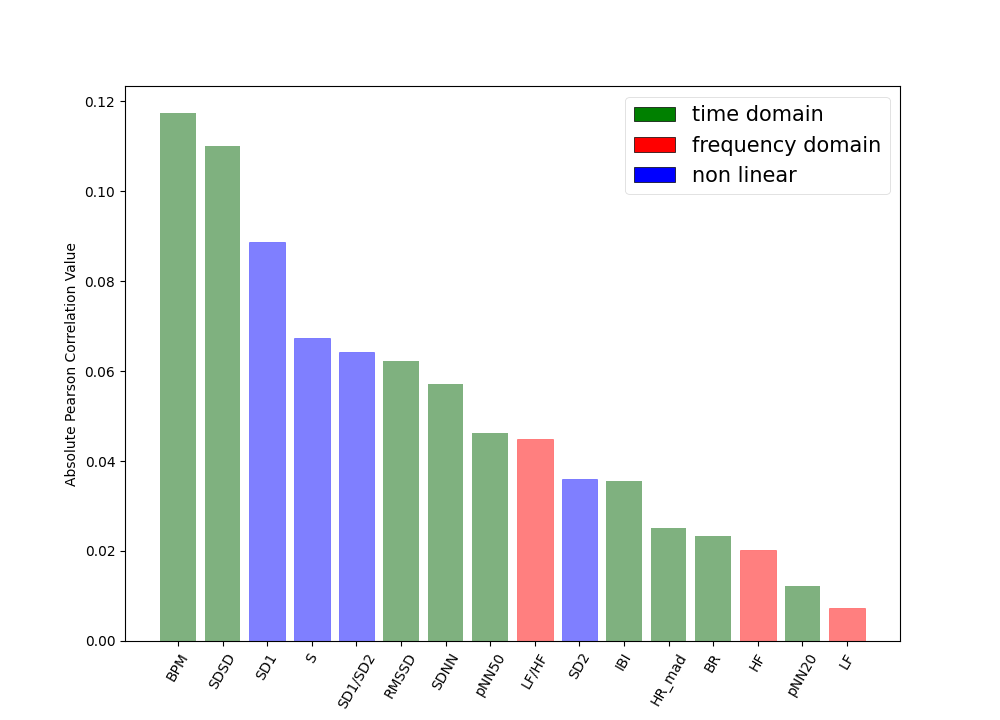}
    \caption{Feature correlations for \textit{Flow} in condition AO.}
    \label{fig:corrfeat_a}
\end{subfigure}%
\begin{subfigure}{.49\textwidth}
	\centering
	\includegraphics[width=.99\linewidth]{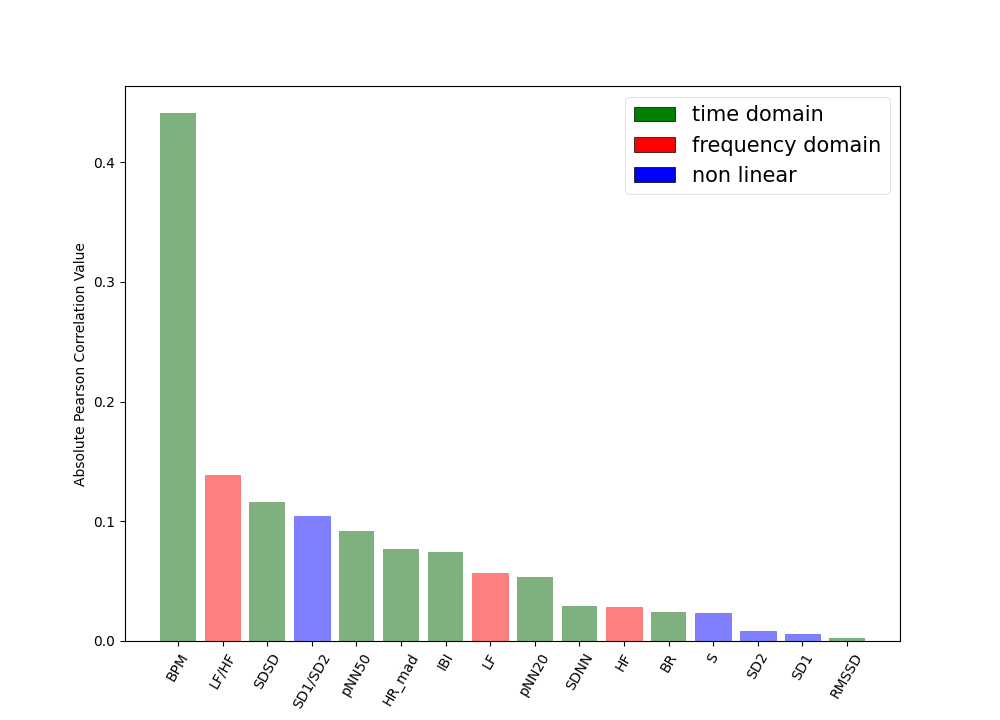}
    \caption{Feature correlations for \textit{Flow} in condition MR.}
    \label{fig:corrfeat_b}
\end{subfigure}
\begin{subfigure}{.49\textwidth}
    \centering
    \includegraphics[width=0.99\textwidth]{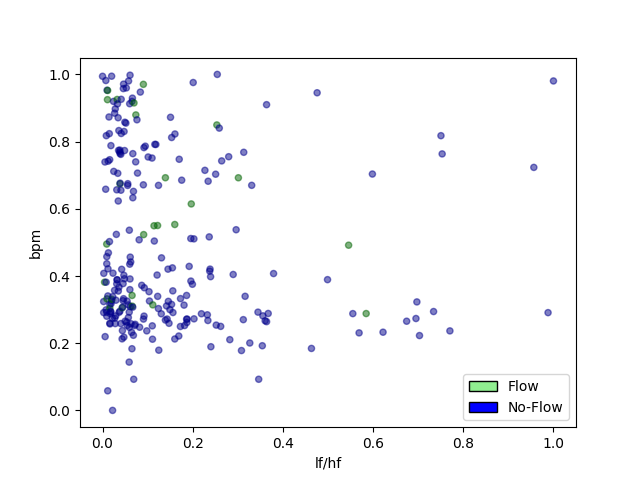}
    \caption{LF/HF and BPM data points for \textit{Flow/No-Flow} in AO.}
    \label{fig:phys_dist_a}
\end{subfigure}
\begin{subfigure}{.49\textwidth}
	\centering
	\includegraphics[width=.99\linewidth]{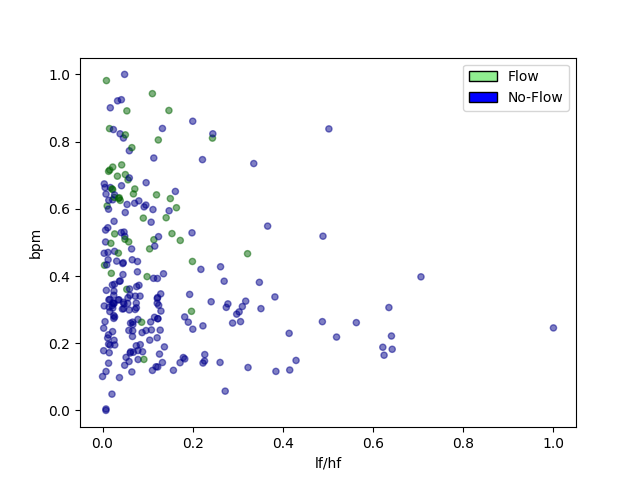}
    \caption{LF/HF and BPM data points for \textit{Flow/No-Flow} in MR.}
    \label{fig:phys_dist_b}
\end{subfigure}
\caption{Results of physiological data analysis for conditions \textit{AudioOnly} (AO, left side) and \textit{MixedReality} (MR, right side).}
\Description{Four subfigures show the results of the physiological analysis. Subfigures 6a and 6b show feature correlations values ranked by value in bar charts. Comparing both subfigures, BPM shows a substantially larger correlation in the MR condition. Subfigures 6c and 6d show data point clusters for flow and no-flow, both in the MR and AO conditions. As dimensions, LF/HF and BPM were chosen for both subfigures. Comparing both subfigures, the data points for flow and no-flow appear to be better separable in the MR condition.}
\label{fig:phys_results}
\end{figure*}

After extracting features, we estimated their discriminative power for differentiating the annotated \textit{Flow} and \textit{No-Flow} time frames. For this purpose, we used the \textit{Correlation Attribute Evaluation} method from the Weka toolkit \cite{witten2005practical}. It analyses the power of a feature by computing the absolute value of the Pearson correlation between it and the class label (\textit{Flow} or \textit{No-Flow}). Figure \ref{fig:phys_results} shows these correlations for condition \textit{AudioOnly} (see Figure \ref{fig:corrfeat_a}) and \textit{MixedReality} (see Figure \ref{fig:corrfeat_b}). As can be seen in both figures, BPM has the highest correlation with the flow label {for both conditions}. However, for the MR condition, the correlation is substantially higher than in the AO condition (Pearson's r = 0.44). LF/HF is the most highly correlated feature from the frequency domainfor both conditions. Within the non-linear domain, SD1 is the most correlated feature for condition AO and SD1/SD2 is the most correlated feature in condition MR. Estimated breathing rate (BR) is also among the features with higher correlation and has the fourth best value. As BPM and LF/HF ratio are the most distinctive features in their respective domains (time and frequency), we plotted the individual data points for \textit{Flow and \textit{No-Flow}} labels in Figures \ref{fig:phys_dist_a} and \ref{fig:phys_dist_b}.

As can be seen in Figure \ref{fig:phys_dist_b}, \textit{Flow} data points within condition MR are generally clustered in lower LF/HF and higher BPM areas as compared to \textit{No-Flow} data points. Lower LF/HF is generally associated with relaxation states and increased PNS activity \cite{shaffer2017overview}, especially under controlled situations such as our laboratory experiment. While higher BPM values for \textit{Flow} in MR (see Figure \ref{fig:phys_dist_b}) are aligned with findings from literature that investigated physiology during music-related flow \cite{de2010psychophysiology}, we did not observe a notable difference between \textit{Flow} and \textit{No-Flow} in the AO condition (see Figure \ref{fig:phys_dist_a}). To quantize the distribution similarities in figures \ref{fig:phys_dist_a} and \ref{fig:phys_dist_b}, we calculated sliced Wasserstein distance \cite{bonneel2015sliced} for both conditions. This metric from transportation theory can be described as the minimum cost to turn one cluster into another. As such, if the metric is higher, the clusters are more separable, e.g., for machine learning applications. When we applied this metric for both conditions, we obtained a value of 11.20 for condition AO and 23.41 for MR. Hence, these results show that the physiological data in Flow and No-Flow states are more separable in the MR condition.


\subsection{Qualitative Data (RQ6 \& RQ7)}
\label{subsec:qualitativeresults}

\begin{figure*}[h!]
\begin{subfigure}{.49\textwidth}
    \centering
    \includegraphics[width=0.99\textwidth]{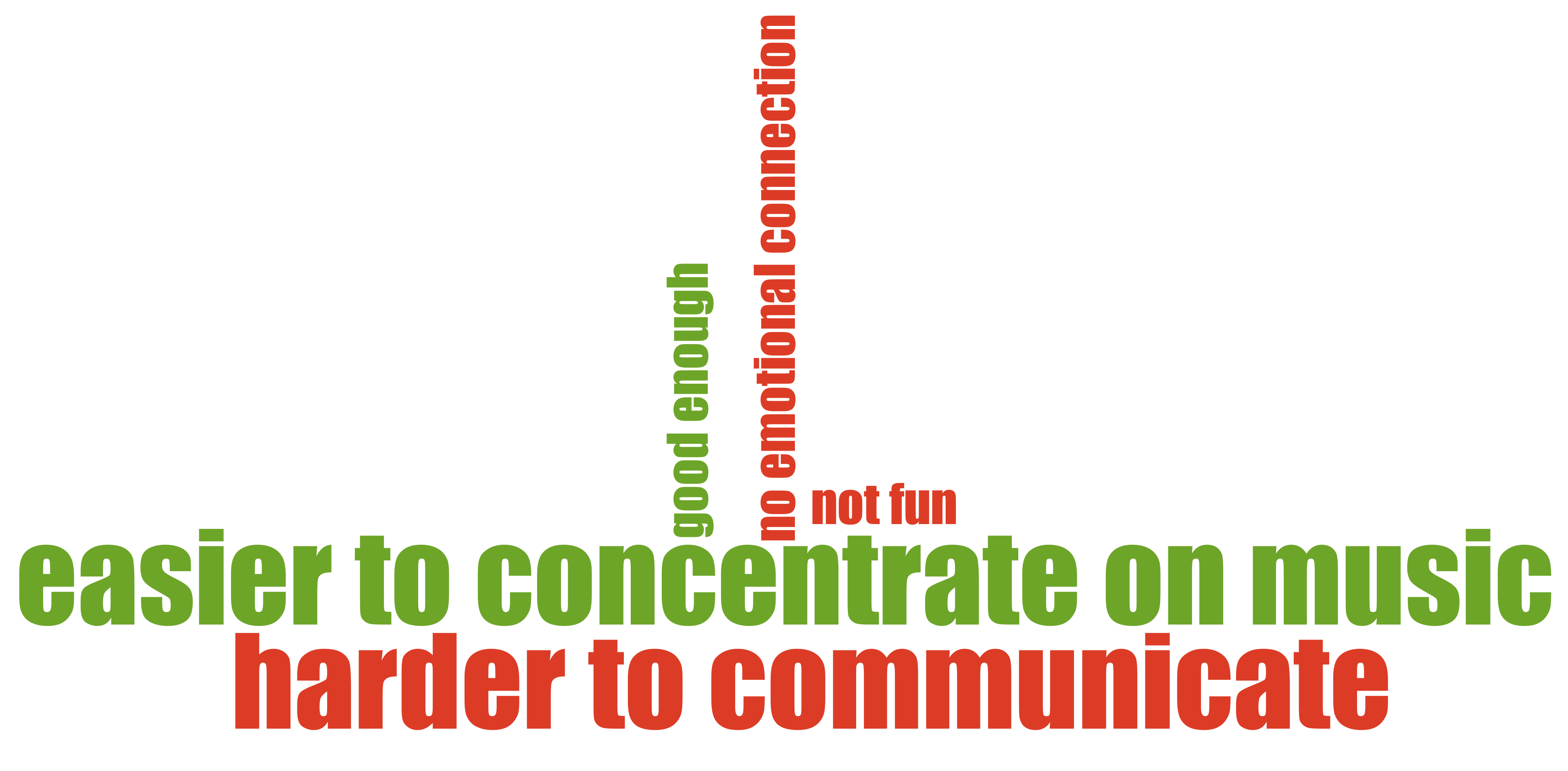}
    \caption{Positive and negative feedback regarding the audio-only jam.}
    \Description{This subfigure shows a code cloud in green and red, showing codes of positive and negative feedback participants gave for the audio-only jam session. The largest is "easier to concentrate on music" for green or positive feedback and "harder to communicate" for red or negative feedback.}
    \label{fig:audio_only}
\end{subfigure}%
\begin{subfigure}{.49\textwidth}
    \centering
    \includegraphics[width=0.7\textwidth]{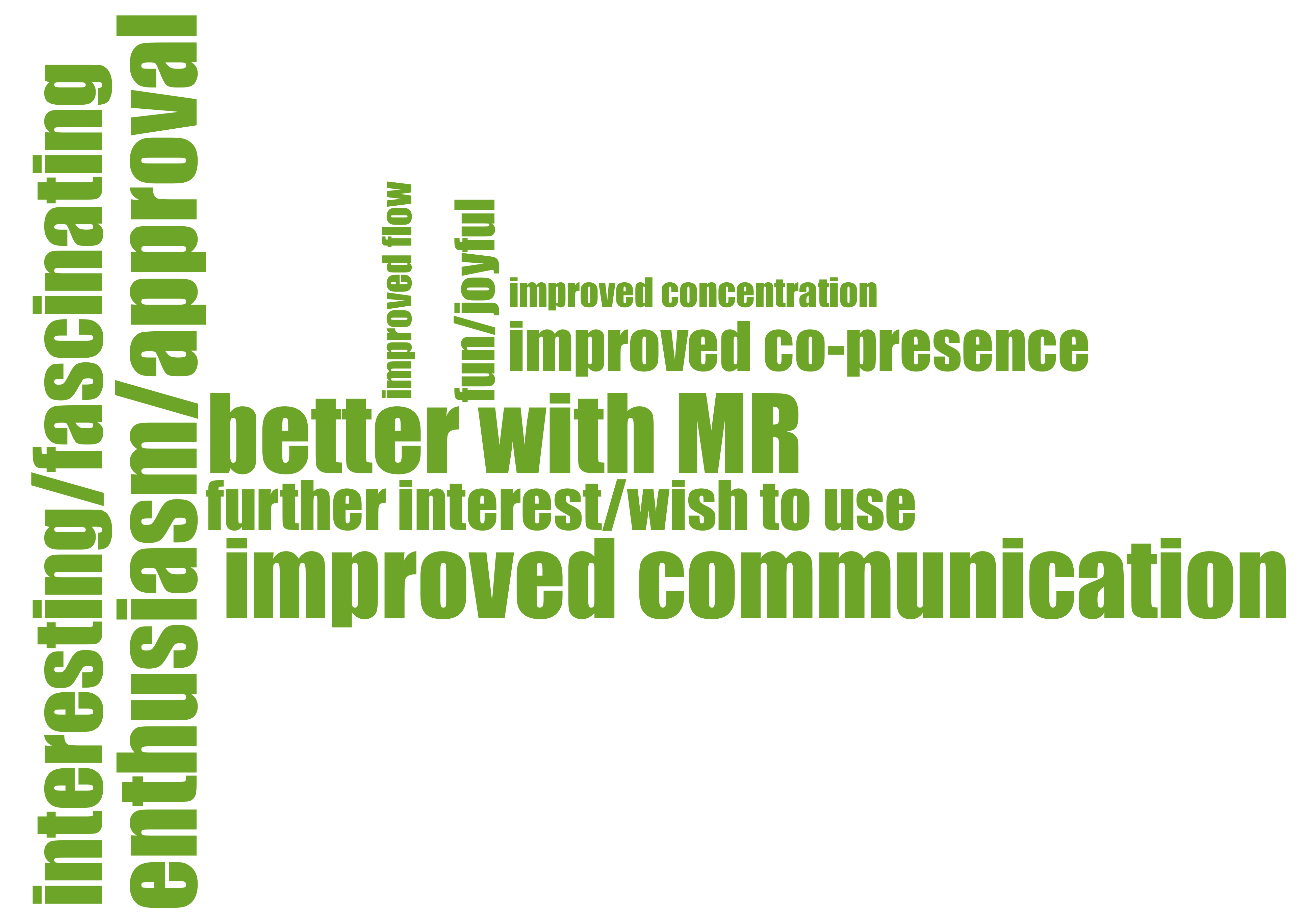}
    \caption{Positive feedback regarding the MR system jam.}
    \Description{This subfigure shows a code cloud in green, showing codes of positive feedback participants gave for the mixed reality jam. The largest codes are "improved communication," "enthusiasm/approval," and "better with MR."}
    \label{fig:mr_positive}
\end{subfigure}
\begin{subfigure}{.49\textwidth}
    \centering
    \includegraphics[width=0.8\textwidth]{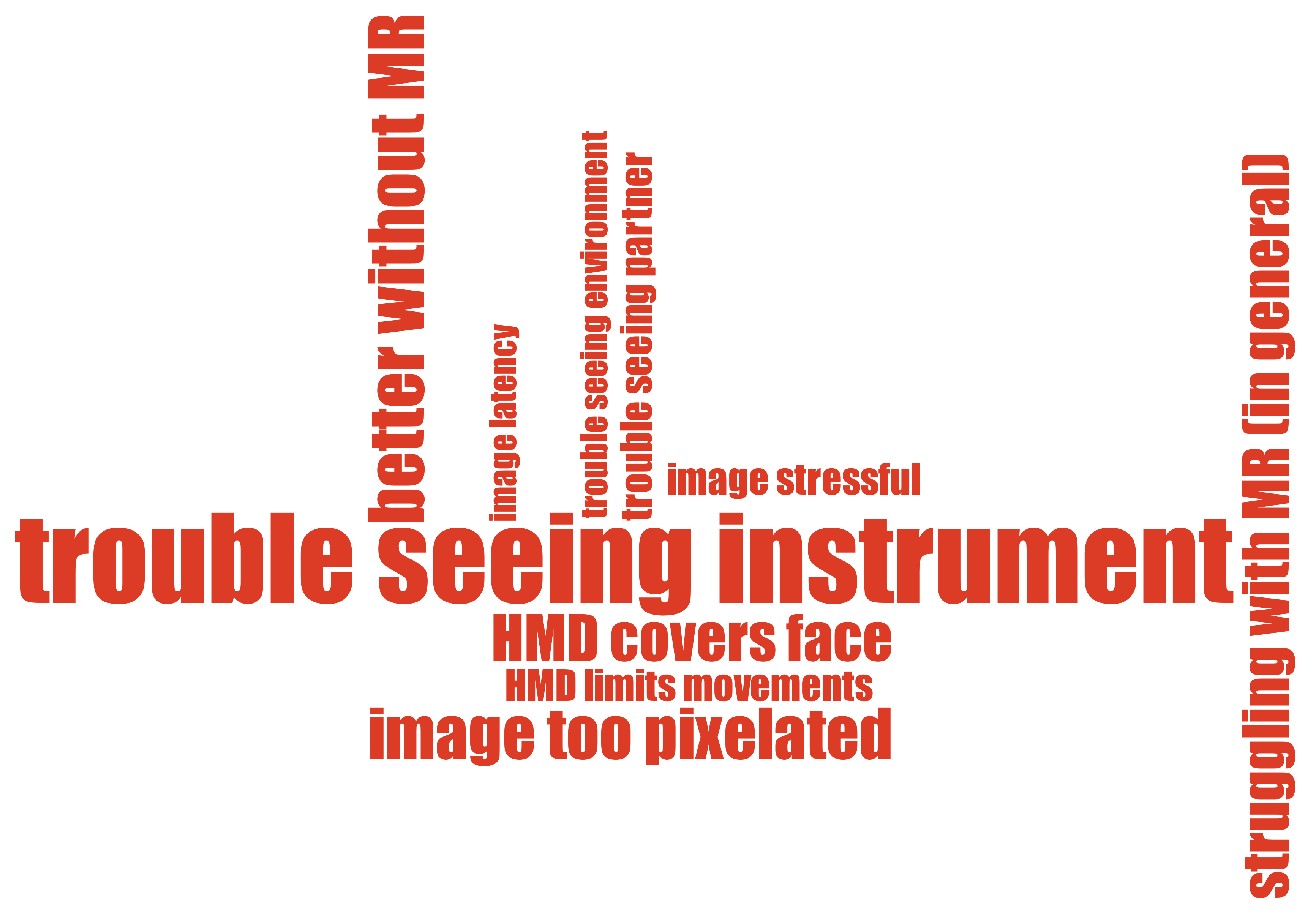}
    \caption{Negative feedback regarding the MR system jam.}
    \Description{This subfigure shows a code cloud in red, showing codes of negative feedback participants gave for the mixed reality jam. The largest codes are "trouble seeing instrument," "better without MR," and "image too pixelated."}
    \label{fig:mr_negative}
\end{subfigure}%
\begin{subfigure}{.49\textwidth}
    \centering
    \includegraphics[width=0.9\textwidth]{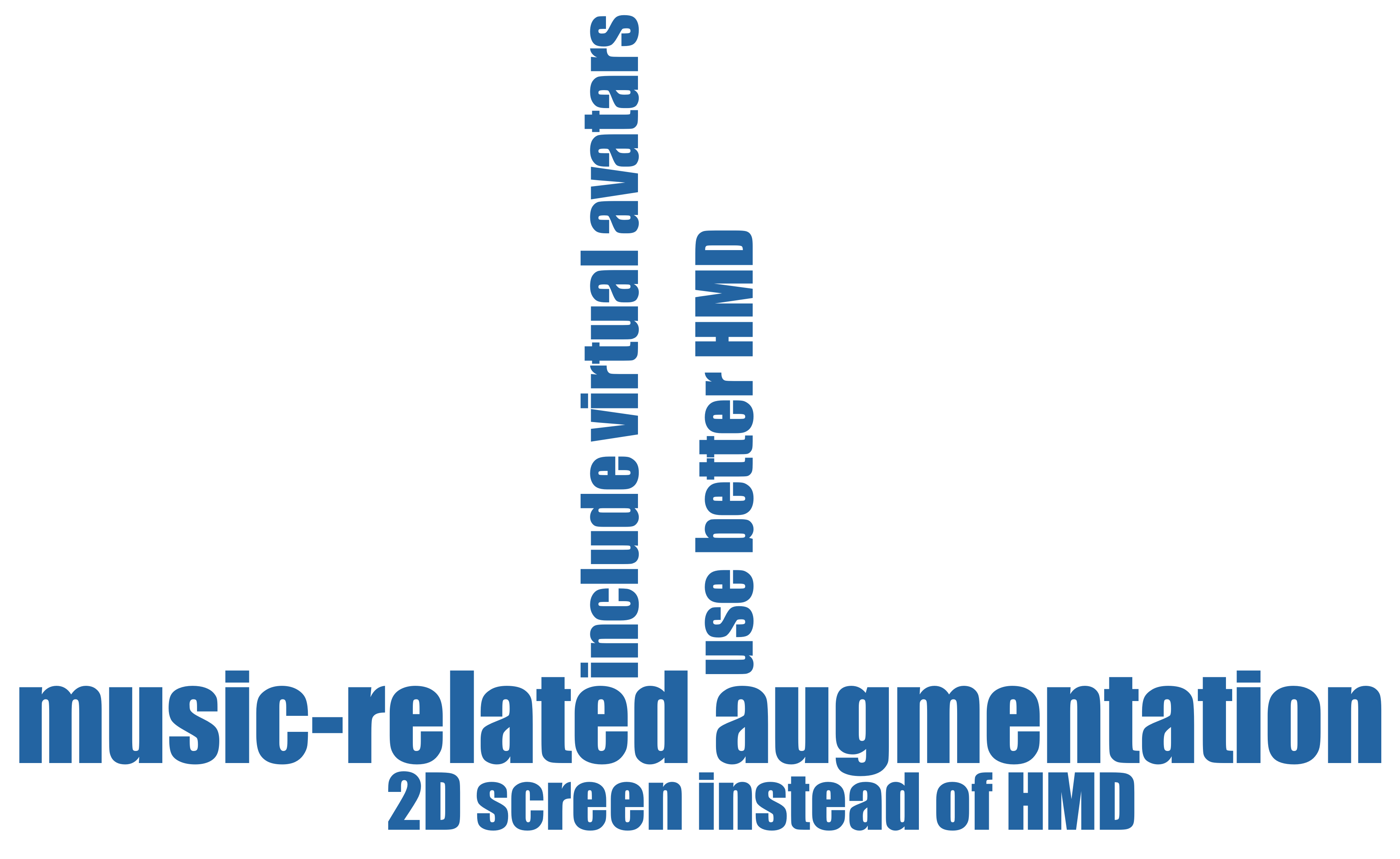}
    \caption{The participants' ideas for improving the MR system}
    \Description{This subfigure shows a code cloud in blue, showing codes of ideas participants uttered for improvement of the mixed reality system. The largest code is "music-related augmentation."}
    \label{fig:mr_ideas}
\end{subfigure}
\caption{Word clouds/code clouds generated from open feedback.}
\label{fig:qual_total}
\end{figure*}

To analyze the written free-form feedback that participants provided after both jam sessions under both conditions, we conducted an inductive thematic analysis \cite{braun2012thematic} using the MaxQDA software\footnote{\url{https://maxqda.com/}}. MaxQDA has been used in various other research projects in HCI (e.g. \cite{krauss2021current, zhang2020enabling, diethei2021sharing}) and allows researchers to code and analyze free-form text such as interview transcripts or, in our case, open feedback that users provided at the end of the experiment. Codes (i.e., category labels) were derived by highlighting important phrases in the participants' answers and summarizing their semantic content in a short descriptive text. After coding the feedback, we determined the frequency of code mentions and created word clouds/code clouds (see Figure \ref{fig:qual_total}) depicting the topics mentioned by participants associated with the qualitative research questions RQ5 and RQ6. 

Figure \ref{fig:audio_only} shows the aggregated positive (green) and negative (red) feedback on the audio-only jam sessions. 
Four participants stated that the audio-only setup made it easier to concentrate on the essence of the jam - the music. On the other hand, an equal number of participants stated that the communication between jamming partners was more difficult than in the MR condition. 
One participant said they found the audio-only setup good enough for jamming, while another complained about a lack of emotional connection with the jamming partner and found an audio-only jam session not fun. 

Figure \ref{fig:mr_positive} shows the aggregated positive feedback on the jam sessions which involved the MR system. Six participants preferred the MR jam session to the audio-only setup. The main benefit of the MR setup appears to be improved communication between the jamming partners, as mentioned by six participants, which mirrors the feedback on the audio-only sessions. Participants also reported improved concentration (one mention), flow (one mention), and co-presence (three mentions) with the MR setup as opposed to the audio-only one. Six participants uttered general enthusiasm and approval for the MR system and five mentioned that they found it to be interesting or fascinating. Three participants wished to use it again or reported having fun and feeling joyful after interacting with the system.

Figure \ref{fig:mr_negative} shows the aggregated negative feedback on the jam sessions which involved the MR system. As seen in the code cloud, the most common point of critique (15 mentions) was the difficulties participants experienced when looking at the instruments to coordinate their hand movements. As the video stream of the user's environment featured a slight but noticeable latency and a substantial screendoor effect, the placement of their hands on their instruments was reportedly sometimes off, causing them to play the wrong notes. Other complaints included poor image quality due to low resolution (seven mentions), relatively high latency (two mentions), and limited motion range because of the HMD (three mentions). Six participants found it difficult to follow their jamming partners' emotional responses during the jam because the HMD obstructed the upper half of their faces. Six participants mentioned struggling with MR in general, either due to technical limitations or because the concept of a video stream perceived through an HMD was perceived to be annoying. In total, eight participants stated that they liked the audio-only jam better. Two participants agreed that this was due to fully concentrating on the music without significant visual distractions. 

Figure \ref{fig:mr_ideas} shows the ideas uttered by some of the participants to improve the MR system. Two participants suggested ideas related to using the HMDs to augment the user's view with music-related information. One suggested augmenting the users' view with notes or guitar tabs for musical pieces. A more complex variant of this idea was uttered by another participant, who would involve a trained AI recognizing the note or chord played by the jamming partner in real-time so that the information could be displayed using the HMD. Another participant thought switching to 2D displays could benefit the MR system. Other ideas suggested switching to an advanced mixed reality HMD featuring improved image resolution (one mention), or implementing virtual avatars instead of the 3D point clouds (one mention). 

\section{Discussion}

\indent\textbf{\textit{Mixed reality point clouds benefit co-presence}}

Comparing the \textit{AudioOnly} and \textit{MixedReality} conditions, we found a significant increase in co-presence, or the feeling of "being there" with the jamming partner. However, since our sample size is relatively small (N=24 after two removed samples), there are limitations regarding the certainty of this observation. Looking at other subscales of the NMSPI questionnaire, there is also a substantial increase in the dimension perceived comprehension (see Figure \ref{fig:social_presence_self_plot}). This is also reflected in the qualitative feedback, where four participants stated that it was harder to communicate in the audio-only jamming situation (see Figure \ref{fig:audio_only}), and six participants noticed it as a strength of the mixed reality setup (see Figure \ref{fig:mr_positive}). We attribute this improvement in perceived communication to the transmission and display of realistic body movements, such as posture and gesture, as the obstruction of faces through the HMDs prohibited communication through eye contact or facial expressions. These findings are in line with previous work that expects realistic embodied avatars and motion to benefit co-presence and social presence (see Subsection \ref{subsec:embodieda_vatars}). However, as our baseline condition did not include visual modalities (as is the case for state-of-the-art remote jamming solutions), it remains unclear whether the increase in co-presence can be attributed to having additional visual modalities in general or to the use of mixed reality technology in particular. Still, we regard the observed increase in co-presence in MR compared to audio-only as a motivation for future developers and researchers to consider MR technologies as a promising approach for including visual modalities in real-time jamming applications, as opposed to just minimizing audio latency. Fostering co-presence may especially be of interest in situations of forced social isolation for musicians, as was the case during the COVID-19 pandemic. In such instances, when meeting in person is not an option, fostering co-presence through MR technologies may overshadow drawbacks associated with technical limitations of current hard- and software, which are discussed later in this section. However, a comparison of our HMD-based MR approach with other display technologies, such as 3D projections or conventional 2D screens, remains to be investigated in the future.

\textbf{\textit{Sometimes less (modalities) is more}} 

Overall, the reception to the MR system was mixed. While six participants stated that they preferred the setup to the audio-only setup in the open feedback (see Figure \ref{fig:mr_positive}), eight enjoyed the audio-only jam more (see Figure \ref{fig:mr_negative}). The main benefit of the audio-only setup was, according to multiple musicians, the increased attention "to the music itself" without any visual distractions. As the point clouds and audio channels did not share the same codes or modalities, we do not attribute this effect to a decrease in cognitive workload \cite{wickens2008multiple}. Instead, we explain this effect with (1) the fascination of seeing real-time 3D point clouds for the first time, which participants reflected in general enthusiasm and fascination (see Figure \ref{fig:mr_positive}) and (2) problems with the video-see through stream latency and resolution (see Figure \ref{fig:mr_negative}), both resulting in increased attention for the visual stimuli. However, as NMSPI scores do not show a decrease in attentional engagement for the musical partner as a whole in the \textit{MixedReality} condition (see Figure \ref{fig:social_presence_self_plot}), we assume the latter to be no significant distraction from the point clouds and audio. Instead, the attention seems to be split mainly between the modalities that were used for communication (audio and visual point clouds). 

\textbf{\textit{Flow comes from the act of remote jamming, whether in mixed reality or not}}

As shown in Figure \ref{fig:flow_level_plot}, musicians in both conditions scored high in their flow rating. The same holds for both split factors, "fluency of performance" and "absorption by activity". While several studies have shown that musicians can enter a flow state while making music, to our knowledge this effect has not been confirmed for remote music making prior to our study. Further, musicians felt challenged and scored low in anxiety, which was found to correlate negatively with flow \cite{kirchner2008relationship}. The high challenge score can be explained by the fact that the participants were not used to jamming in remote situations, regardless of the condition. Also, the technical limitations of the MR system might play a role (they are discussed in further detail below). As we did not see an increase in any flow dimension for the MR condition, flow and the feeling of co-presence in MR do not seem to be correlated. We assume that the ability of some participants to focus on the audio exclusively and experience increased flow in this concentrated state might have outweighed any immersive effects caused by the mixed reality system. Since a basic requirement for flow is to have challenges that are well suited to individual skills \cite{csikszentmihayli1991talent}, this is especially expected to be the case for experienced musicians, which we can confirm from our observations.
The low anxiety score, however, is unexpected since the test leaders recorded the audio and listened to the jam in the control room. However, the free choice of the respective jamming partners might have decreased the participants' performance anxiety and thus benefited flow \cite{kirchner2008relationship}. 

\textbf{\textit{Remote jamming increased positive affect}}

Regardless of the condition, we measured a significant increase in positive affect after the first remote jam session. However, we did not observe any significant effect between the \textit{AudioOnly} and \textit{MixedReality} conditions. In order to investigate effects on well-being, long-term studies would need to be conducted that consider positive and negative affect independently \cite{diener1984independence}. As the mixed reality system significantly increased co-presence, an effect on positive affect might be observable if musicians would otherwise be affected by loneliness or isolation. Therefore, we might not have measured the true potential of increasing positive affect by fostering co-presence, as we conducted the experiment when no social distancing measures were active. However, as these measures were dropped not long ago prior to the experiment, many musicians reported that they had not jammed with a musical partner for months or even years. As such, musicians might have been happy to jam after a long time which might have increased the measured pre- vs. post-first-jam effect, overshadowing a condition-based effect behind in terms of effect size.

\textbf{\textit{Technical limitations are a key issue}}

The most prominent points of critique uttered by participants in the qualitative feedback were related to the technical specifications of the head-mounted displays that were in use (HTC Vive Pro models) and their video-see-through mode. In particular, the resolution and noticeable delays of the video streams of the musicians' surroundings were causing problems, especially regarding precise finger- and hand placements on their own instruments. These issues even caused some (especially older) participants to reportedly close their eyes during the MR jam to focus solely on the audio modality. These issues could significantly be decreased by using better state-of-the-art HMDs such as the XR-3 model by Varjo\footnote{\url{https://varjo.com/products/xr-3/}}. Another option to remove pass-through video artifacts is the use of optical see-through hardware. However, state-of-the-art headsets like the Microsoft HoloLens 2\footnote{\url{https://microsoft.com/hololens/}} have significant drawbacks regarding the field of view and occlusion. Another option is to move away from HMD-based AR entirely and use perspective-corrected projections such as in \cite{pejsa2016room2room}. 
This approach could provide a better resolution and unobstructed faces. However, as of now, it is unclear whether this approach might be comparable regarding experienced co-presence. Future studies could compare this approach to our HMD-based setup.
Interestingly, few participants complained about the resolution of the point clouds, which could be improved by incorporating more or better depth cameras, and no participant complained about point cloud latency. However, the laboratory setup we used bypasses some latency issues by using an analog audio setup and the local network for point cloud transmission. Due to this circumstance and the need for specialized hardware, our system only provides isolated jamming capabilities for musicians willing to visit the lab. For low bandwidth remote jamming, e.g., via the internet, we propose using motion-tracked virtual avatars, as they only require the transmission of joint poses.
Another problem is the obtrusion of the upper face region by the HMDs, prohibiting communication through eye fixations or facial expressions, which have been found to increase the amount of timing synchrony during jam sessions \cite{morgan2015using}. This circumstance is also reflected in the NMSPI dimension "emotional contagion", that even shows a downward trend for the \textit{MixedReality} condition (see Figure \ref{fig:social_presence_self_plot}). As such, we highly encourage endeavors for tracking and reconstructing facial movement for similar MR systems.

\textbf{\textit{Latency remains a constraint for remote jamming}}

A key challenge that needs to be tackled to allow for real-time music collaboration over larger distances remains signal transmission latency. While solutions to this problem are actively researched in the field of \textit{Networked Music Performance} \cite{rottondi2016overview}, we bypassed this challenge in our experiment by using analog audio hardware to locally connect two laboratory rooms. For signal transmission via the internet, digital protocols such as UDP need to be used to reduce signal loss which increases significantly with distance when using analog signals. Since the speed of light is a theoretical limit for communication speed over the internet, network latency is at least one additional millisecond per 300 kilometers of the signal path \cite{drioli2013networked}. Audio latency in real-time audio monitoring is considered "fair" in a range between 3-43 ms, depending on the instruments used and monitoring system \cite{lester2007effects}. However, in practice, latency requirements also depend on various aspects such as reverberation effects, performed tempo, proficiency of musicians, and more \cite{rottondi2016overview}. Taking the 43 ms of study \cite{lester2007effects} as an upper threshold, 12.900 km of the overall signal path, which is about 32\% of the earth's equator length and 64\% of the maximum distance on earth between two musicians, remains a physical limit for one-way communication with "fair" latency. Since the response of a remote musician to a note played by the other musician implies a two-way path, the maximum distance is halved to approximately 32\% of the theoretical maximum distance between two musicians on planet earth. Since additional delays (e.g., analog-to-digital conversion (ADC) and digital-to-analog conversion (DAC), among others \cite{rottondi2016overview}) also need to be considered for both sides, this distance is even shorter in practice. To conclude, real-time remote jamming applications that suffice musicians' needs for latency are unlikely to be solved for arbitrarily long distances between musicians. However, better protocols and higher bandwidth may increase the spatial distance for which real-time remote jamming may be feasible, and we encourage further research in this area.

\textbf{\textit{Flow has distinctive features that may enable real-time assessment}}

A fundamental problem when measuring flow is that assessment methods such as self-report questionnaires will most likely interrupt the act in which test subjects are involved and disrupt their flow state. As such, real-time assessment of the flow state is typically not practical. Analyzing the physiological data we gathered during  jam sessions, we found EDA features that were descriptive for the state of flow, namely heart rate (BPM) and HF/LF while musicians were jamming in mixed reality. While flow-induced increases in BPM are known from previous work \cite{de2010psychophysiology}, the decrease in HF/LF was not yet correlated with flow. Instead, it is generally associated with relaxation states, and increased PNS activity \cite{shaffer2017overview}. We assume that this might be a feature that works specifically for the jamming context, where flow is a goal or common sense. Accordingly, anxiety might decrease as soon as participants enter the flow state. As the observed increase in heart rate is consistent with the literature, we interpret this as evidence for the validity of our post-stimulus self-annotation methodology. However, during these annotation sessions, we experienced some misunderstandings that participants had, confusing the state of flow, which is the act of being fully immersed in the act of making music and \textit{musical attunement}, which jazz musicians refer to as `striking a groove' \cite{seddon2005modes}. While a correlation between these two phenomena seems intuitive, it is not yet empirically validated to our knowledge. To our surprise, the positive correlation between heart rate and flow was particularly high in the \textit{MixedReality} condition (see Figure \ref{fig:corrfeat_b}). An explanation for this phenomenon might be that musicians were more physically demanded in intense improvisation moments in MR, as they wore HMDs while simultaneously intensifying their musical performances. In addition, the higher BPM correlation seems to have contributed to the fact that the clusters of \textit{Flow} and \textit{No-Flow} data points were particularly well separable within the \textit{MixedReality} condition (see Figure \ref{fig:phys_dist_b}).

Based on our findings and drawing inspiration from previous work that used VR to foster flow \cite{person2018flow,ruvimova2020transport}, we propose the idea of using classifiers such as support vector machines or artificial neural networks for real-time flow assessment, which could then serve as an input for a mixed reality system that induces flow by providing extra stimuli for musicians. For instance, extra assistance could be provided, such as determining the musical key for over-challenged amateur musicians. Similarly, extra challenges, such as displaying off-key notes that need to be incorporated, could be included for expert musicians to provide an extra challenge.

\section{Conclusion}

This paper reported on a laboratory study that examined the effect of remote jamming in mixed reality on co-presence, affect, and flow. Furthermore, we looked at physiological signals to determine features that are distinct for the state of flow during remote jam sessions. While we did not observe a notable difference between the baseline and mixed reality condition regarding positive affect or flow, we found a significant increase in the participants' feeling of co-presence while jamming in MR. Furthermore, we found that the act of remote real-time music collaboration, regardless of condition, significantly increased the musicians' positive affect. 
From the physiological signal analysis, we determined heart rate and HF/LF as promising features for classifying the self-reported flow state musicians experienced while making music together.
In our discussion, we contextualized these results with the findings from the qualitative feedback and proposed ideas for future mixed reality systems that might benefit musicians while jamming remotely in real-time. Overall, we conclude that mixed reality can benefit remote music collaboration, especially in social distancing scenarios, as it benefits the feeling of co-presence. However, besides giving musicians a sense of "being there" with their musical partners, MR technologies offer a large space for creative design solutions that might improve the experience of real-time remote music collaboration in other ways. One promising approach that our results encourage is to develop affective computing systems that consider and respond to the musicians' flow state, which could be assessed in real-time using physiological data.

\begin{acks}
This paper was partially funded by the DFG through the Leibniz award of Elisabeth André (AN 559/10-1).
\end{acks}

\bibliographystyle{ACM-Reference-Format}
\bibliography{main}



\end{document}